\documentstyle[12pt]{article}
\def\be{\begin{equation}}
\def\ee{\end{equation}}

\def\simlt{\lower.5ex\hbox{\ltsima}}
\def\gtsima{$\; \buildrel > \over \sim \;$}
\def\simgt{\lower.5ex\hbox{\gtsima}}
\def\hmpc{{\rm\,h^{-1} Mpc}}
\def\msun{{\rm\,M_\odot}}
\def\msunh{{\rm\,h^{-1}~M_\odot}}

\def\ergcm2{\ {\rm erg~cm^{-2} }}
\def\ergscm2{\ {\rm erg~s^{-1}~cm^{-2} }}
\def\hmpc{\ h^{-1}~{\rm Mpc}}

\def\s{\ifmmode \widetilde \else \~\fi}
\def\={\overline}

\def\spose#1{\hbox to 0pt{#1\hss}}

\def\eg{{\it e.g.\ }}
\def\etal{{\it et al.\ }}
\def\ie{{\it i.e.\ }}

\def\lta{\mathrel{\spose{\lower 3pt\hbox{$\mathchar"218$}}
     \raise 2.0pt\hbox{$\mathchar"13C$}}}
\def\gta{\mathrel{\spose{\lower 3pt\hbox{$\mathchar"218$}}
     \raise 2.0pt\hbox{$\mathchar"13E$}}}
\def\mincir{\ \raise -2.truept\hbox{\rlap{\hbox{$\sim$}}\raise5.truept  %MC
\hbox{$<$}\ }}                                                          %
\def\magcir{\ \raise -2.truept\hbox{\rlap{\hbox{$\sim$}}\raise5.truept  %
\hbox{$>$}\ }}                                                          %
\def\simlt{\ \raise -2.truept\hbox{\rlap{\hbox{$\sim$}}\raise5.truept   %MC
\hbox{$<$}\ }}                                                          %
\def\simgt{\ \raise -2.truept\hbox{\rlap{\hbox{$\sim$}}\raise5.truept   %
\hbox{$>$}\ }}                                                          %
\def\newline{\smallskip\noindent}

\def\t{{\tau}}

\def\ea{{\it et al.} \,}

\def\s-z{S-Z}
\def\cmb{CMB\,}
\def\comp{Comptonization\,}

\def\pr{$^\prime$}

\begin{document}
%\flushright{\date{today}}
%\baselineskip=24pt
\baselineskip=18pt
%\vspace*{0.3in}

\centerline{\bf INTRACLUSTER COMPTONIZATION OF THE CMB: MEAN SPECTRAL}
\medskip

\centerline{\bf DISTORTION AND CLUSTER NUMBER COUNTS}

%\vspace*{0.3in}
\vskip 0.5truecm

\centerline{S. COLAFRANCESCO}
\vskip 0.1truecm
\centerline{Osservatorio Astronomico di Roma}
\centerline{via dell'Osservatorio, I-00040 Monteporzio, Italy}
\vskip 0.3truecm
\centerline{P. MAZZOTTA}
\vskip 0.1truecm
\centerline{Dipartimento di Fisica,  Universit\`a di Roma ``Tor Vergata''}
\centerline{via della Ricerca Scientifica 1, I-00133 Roma, Italy}
\vskip 0.3truecm
\centerline{Y. REPHAELI$^{\star}$}
\vskip 0.1truecm
\centerline{Center for Particle Astrophysics}
\centerline{University of California, Berkeley, CA 94720}
\vskip 0.3truecm
\centerline{Physics Department}
\centerline{Stanford University, Stanford, CA 94305}
\vskip 0.3truecm
\centerline{and}
\vskip 0.3truecm
\centerline{N. VITTORIO}
\vskip 0.1truecm
\centerline{Dipartimento di Fisica,  Universit\`a di Roma ``Tor Vergata''}
\centerline{via della Ricerca Scientifica 1, I-00133 Roma, Italy}

\vskip 0.5truecm
%Submitted for publication in \underline{\it The Astrophysical Journal}
%}
%\vspace*{0.5in}
\vskip 0.5truecm
\noindent
{\it Subject headings:} cosmic microwave background --
cosmology: theory -- clusters of galaxies: evolution.
\smallskip\noindent
%{\it Subject headings:} ~cosmology: background ~radiation ~--
%~galaxies: ~clustering ~-- ~background: ~radiation.

%\vspace*{0.3in}
\vskip 0.3truecm
\noindent
$^{\star}${On leave from School of Physics and Astronomy, Tel Aviv
University, Tel Aviv, Israel}
\newpage
%\baselineskip=16pt
%\vspace*{2.0in}

\centerline{\Large\bf Abstract}
\noindent
The mean sky-averaged \comp parameter, $\overline{y}$, describing the
scattering of the \cmb by hot gas in clusters of galaxies is calculated
in an array of flat and open cosmological and dark matter models. The models
are globally normalized to fit cluster X-ray data, and intracluster gas
is assumed to have evolved in a manner consistent with current observations.
We predict values of $\overline{y}$ lower than the COBE/FIRAS upper limit.
The corresponding values of the overall optical
thickness to Compton scattering are $\simlt 10^{-4}$ for relevant parameter
values. Of more practical importance are number counts of clusters
across which a net flux (with respect to the CMB) higher than some limiting
value can be detected. Such number counts are specifically predicted
for the COBRAS/SAMBA and BOOMERANG  missions.
% (which are currently under study).

%\end{document}
\baselineskip=24pt

\vskip 0.3truecm
\noindent
{\Large\bf 1 \, Introduction}

\noindent
Compton scattering of the cosmic microwave background (CMB) radiation by
hot gas in clusters of galaxies -- the Sunyaev-Zeldovich (1970; hereafter
S-Z) effect -- affects both the spectrum and the angular distribution of
the radiation. The spectral change of the CMB across each cluster translates
into a superposed average change of the spectrum across the sky.
When the intensity (or temperature) is differentially measured, the
radiation will appear anisotropic, on characteristic angular scales of
several arcminutes, reflecting the typical sizes of clusters that
contribute most to the anisotropy.
This cluster-induced anisotropy is an important component of the full
anisotropy on such scales, as has been shown in many studies (see, \eg,
Rephaeli 1995a, and references therein).

Less studied is the overall spectral effect of clusters on the CMB.
Whereas the anisotropy is a measure of the rms value of the \comp
parameter $y$ [defined in Equation (12) below], the overall spectral effect
results from the superposed effects of individual clusters
over the sky. This parameter is of prime interest in the characterization
of spectral deviations from a pure blackbody spectrum. Currently, the
COBE/FIRAS database provides the best measure of such non-Planckian
distortions by constraining the degree of \comp to be  $y \leq 1.5\times
10^{-5}$ at the 90\% statistical significance level (Fixsen \ea 1996).
The significance of this value stems from the fact that it limits the
cumulative effects of energy release processes in the early universe
(at $z \simlt 3\times 10^6$), and the superposed effects of hot
intergalactic (IG) and intracluster (IC) gas (Wright \ea 1994).
In order to predict the contribution to the \comp parameter due to a
population of evolving clusters,
some assumptions need to be made about the
evolution of groups and clusters (which, for brevity, will hitherto be
collectively referred to as clusters) of galaxies, and basic
characteristics of their hot gas. Under these assumptions the distribution
of the \comp parameter in clusters at different redshifts and in a given
mass range can be calculated. From this, the mean \comp due to clusters
can be reasonably well estimated.

Here we present the results of an
investigation of the effects on the CMB of a population of evolving
clusters, described in the background of various flat and
open cosmological models.
In a previous paper (Colafrancesco \ea 1994) we have
reported results of a calculation of the predicted rms fluctuations in
the CMB induced by gas in clusters in a set of flat cosmological models.
Because our basic approach here is similar to that detailed in the latter
paper, our discussion -- in $\S 2$ -- of the general background and
modeling of clusters and of their IC gas will be brief. The basic
definitions of the sky-averaged quantities we calculate here are given
in $\S 3$. We also calculate the expected cluster number counts
in the mm and sub-mm bands.

Long duration balloon flights and space missions dedicated to the study of
CMB anisotropy are expected to reach sensitivity levels of
$\approx 10~\mu{\rm K}$ per pixel. The COBRAS/SAMBA mission (hereafter
C/S; see \eg Mandolesi \ea 1995),  has been selected as the
next Medium-Sized Mission of the ESA Horizon 2000
Program. The eight channels of this experiment cover a wide frequency
range, from 35 to 714 GHz. In particular, the 140, 222 and 400 GHz channels
of the High Frequency Instrument (HFI) are particularly relevant for
measurement of the S-Z effect in clusters.
The first frequency is
on the Rayleigh-Jeans (R-J) side, where the intensity change (across a
cluster) is negative; the second is near the crossover frequency (where the
thermal effect vanishes), and the third is on the Wien side, where the
effect is positive (for a discussion of the significance of high-frequency
measurements, see Rephaeli,  1995a). The corresponding angular
resolutions at these frequencies are 10.5\pr, 7.5\pr, 4.5\pr (FWHM),
respectively. A long duration balloon-borne experiment BOOMERANG
(see Lange \ea 1995) is expected to cover a $10^o\times 10^o$ patch
of the sky with
a FWHM resolution of 5\pr at 150 GHz. The projected sensitivity levels and
angular resolution of these experiments are very suitable for a survey
of a large number of clusters. In $\S 4$ we present the results of
our calculations giving the predicted ranges of values of basic quantities
which can be determined in light of these observational capabilities.
Finally, in $\S 5$ we assess these results and summarize our main
conclusions.
\bigskip

\vskip 0.3truecm
\noindent
{\Large\bf 2 \, Cosmological Evolution of Clusters}
\smallskip

\noindent

\vskip 0.2truecm
\centerline {\it 2.1 Mass and Redshift Distribution of Clusters}

\smallskip
We adopt the simple spherical collapse picture for cluster formation,
according to which a homogeneous, spherical perturbation
detaches from the Hubble flow at time $t_m$, collapses at time
$t_c\simeq 2 t_m$, and virializes at time $t_v\simeq 3 t_m$.
The relative density contrast at an initial redshift $z_i$,
$\delta_{i,v}$, depends on the cosmological model.
Under the assumption of linear
growth, the density contrast at $t_v$ is
$\delta_v=\delta_{i,v}D(t_v)/D(t_i)$, where $D(t_v)$ is the linear
growth factor in the chosen cosmology.
The relevant formulae for low density (open, or
vacuum dominated) models are given in the Appendix.
For $\Omega_0\rightarrow 1$, $\delta_v$ tends to the standard value of
$2.2$, independent of $t_v$.
The actual, non-linear density contrast
is $\Delta = {\tilde \rho}/ \rho$, where ${\tilde \rho}$ is the
perturbation density and $\rho$ is the background density at the time
of virialization. In the Appendix we also briefly review how to derive
$\Delta$ in low density cosmological models.
For $z_v \rightarrow \infty$, $\Delta$, evaluate at $z_v$
tends to the standard value
$\approx 400$, found in a flat universe.

The mass and redshift distribution of clusters, $N(M,z)$, can be
determined from the Press \& Schechter (1974; hereafter P\&S) theory:
\begin{equation}
N(M,z) = {\rho\over M} {\delta_v \over \sigma^2}{d\sigma \over dM}
{1\over \sqrt{2\pi}}\exp[-\delta_v^2/2\sigma^2] \, ,
\end{equation}
where $\rho$ is the background density at redshift $z$, $M$ is the
total cluster mass, $\sigma(M,z)$ is the {\it rms} of the linear density
fluctuation field at $z$, smoothed over the region containing $M$,
and $\delta_v$ is the linear density contrast of a perturbation that
virializes at $z$. The variance of density fluctuations of mass $M$
is given in the standard relation:
\be
\sigma^2(R,z) = D^2(z) \int dlnk~ k^3 P(k) \bigg[ {3 j_1(kR) \over kR}
\bigg]^2
\ee
with
$M = 4 \pi \rho R^3 / 3$. We use the fitting formulae for the power
spectra, $P(k)$, given in Holtzman (1989) in order to calculate the
mass variance in different cosmological models. Here we consider
flat cold (CDM) and mixed (MDM) dark matter models, low density CDM
models with a cosmological constant (CDM$+\Lambda$), as well as open,
pure baryonic models with isocurvature initial conditions (BDM).

If the power spectrum of density fluctuations is
a power-law of the spatial frequency [\ie $P(k)=Ak^n$], then
$\sigma(M,z)=(1/b)(M/M_0)^{-\alpha} D(z)$, where
$M_0$ is the mass contained in a sphere of
$8h^{-1}Mpc$ radius, $b$ is the biasing factor, and
$\alpha=(n+3)/6$. For this power-law spectrum, the mass distribution
assumes the well known form:
\begin{equation}
N(M,z)= {{\cal I} \over \sqrt{2\pi}}
{\rho\over M_0^2}{n+3\over 6} {\delta_v b \over D(z)}
\bigg({M\over M_0} \bigg)^{\alpha -2}
\exp\bigg[-{1 \over 2} {\delta_v^2 b^2 \over D^2(z)} \bigg({M\over M_0}
\bigg)^{2\alpha} \bigg] \, ,
\end{equation}
where $1 \leq {\cal I} \leq 2$ takes into account
possible secondary infall of mass into the cluster after
its initial collapse and virialization.

The fact
that $N(M,z)$ depends on the product $\delta_vb$, and not
separately on $\delta_v$ and $b$, somewhat simplifies the fit to
X-ray data. We
have determined the
values of ${\cal I}$ and $\delta_vb$ in various dark matter models
by fitting to the observed cluster X-ray luminosity
function of Kowalski \etal (1984).
The results of our analysis (see Colafrancesco \& Vittorio 1994 for
details) are showns in Table 1.
Using a P\&S theory where ${\cal I}$ and $\delta_v b$ are fitted to
the X-ray luminosity (or temperature) functions, weakens the predictive
power of the theoretical models based on the linear theory. However,
the fitted values of ${\cal I}$ and $\delta_v b$ should contain the
relevant informations of a more realistic cluster formation picture.

\vskip 0.2truecm
\centerline {\it 2.2 Properties and Evolution of Intracluster Gas}

\smallskip
There are various open issues pertaining to the formation of the hot
gaseous cores of clusters. For our purposes here we can adopt the
following simplified approach: shortly after a cluster forms and
virializes, a gaseous core forms (probably as a result of tidal galactic
interactions and other gas stripping processes) with the hot gas in
hydrostatic equilibrium in the potential well of the cluster.
To avoid
introducing a large number of free parameters, we will simply scale
the gas properties to those of the cluster. This is suggested by X-ray
observations of local clusters (see, \eg, Jones \& Forman 1992).
The gas mass is taken to be a fraction $f$ of the total cluster mass,
and the gas density profile is assumed to have the commonly adopted form
\begin{equation}
n(r) = n_0[1+(r/r_c)^2]^{-3\beta/2} \, ,
\label{npro}
\end{equation}
where $n_0$ is the central electron density and $r_c$ is a core radius.
The observed values of $\beta$ range roughly from 0.5 to 0.7. Here we
use the value $\beta=2/3$ which is particularly convenient in
analytic calculations.

The radial extent of a cluster is taken as
$R=pr_c$. The mass within the outer radius $R(p)$ is
\begin{equation}
M(p) = 3 M_0 [p-tg^{-1}p] \, ,
\end{equation}
where $M_0=(4\pi/3)r_c^3\rho_0$, and $\rho_0$ is the central total mass
density of the cluster.

Assuming that cluster collapse is
self-similar, we can infer the mass and redshift dependence of $r_c$ from
the scaling law:
$
r_c = \bigg[3 M / 4 \pi \rho_b \Delta\bigg]^{1/3}(1+z)^{-1}.
$
This gives
\begin{equation}
r_c(\Omega_0,M,z) = {1.29 \hmpc \over p} \biggl[ {M\over 10^{15} \msunh}
\cdot {\Delta(\Omega_0=1,z=0)\over \Omega_0 \Delta(\Omega_0,z)}\biggr]^{1/3}
{1\over 1+z} \, .
\end{equation}
Hereafter we fix $p=10$ to recover values of the IC gas core radii
consistent with  observations
(see, \eg, Sarazin 1988, Jones \& Forman 1991). In particular,
for a local cluster of $10^{15} \msunh$ we get
$r_c=0.12$ and $0.16\hmpc$ for $\Omega_0=1$ and
$\Omega_0=0.2$, respectively.

The gas is assumed isothermal at the
virial temperature $T \propto M/R$. This gives
\begin{equation}
T = 8.7 \cdot 10^7 (1+z)({M \over 10^{15} h^{-1} M_{\odot}})^{2/3} 
\cdot \biggl[ {\Omega_0 \Delta(\Omega_0,z) \over \Delta(\Omega_0=1,z=0)}
\biggr]^{1/3} \,K.
\end{equation}

Although it is known that the gas mass fraction depends on $z$ and
$M$, little is currently known on the exact form of these dependences.
We adopt the simple parametrization (described in detail in
Colafrancesco \& Vittorio 1994; see also Cavaliere, Colafrancesco
and Menci 1993)
which is based on the results of analyses of the {\it Einstein} Medium
Sensitivity Survey (EMSS) data (Gioia \ea 1990; Henry \ea 1992)
which seem to indicate a decrease in the number of bright clusters with
redshift, and an analysis of a local cluster sample (David \ea 1990):
\begin{equation}
f = f_o \biggl({M\over 10^{15} \msunh }\biggr)^\eta
\bigg({t\over t_o}\bigg)^\xi \, .
\end{equation}
Here $t_o$ is the age of the universe, and the normalization to
$f_o \simeq 0.1 $, is based on a local, rich cluster sample.
Values of $\eta$ and  $\xi$ are listed in Table 1.
\bigskip

\vskip 0.3truecm
\noindent
{\Large\bf 3 \, Cluster \comp}
\smallskip

\noindent
In the non-relativistic limit, the effect of scattering of the CMB by
hot gas depends linearly on the cluster \comp parameter:
\begin{equation}
y_c = {kT\over m_ec^2} \sigma_T \int_{\ell_{min}}^{\ell_{max}} n d\ell \, ,
\end{equation}
where $\sigma_T$ is the Thomson cross section, and the integral is over a
line of sight through the cluster. In the exact relativistic treatment, the
dependence of the effect is not exactly linear in $y$. For
low values of the optical thickness $\t$ to Compton scattering, the
dependence on $\t$ is linear, but the dependence on $T$ is more
complicated, even
for the observed range of IC gas temperatures, roughly $3 \div 15$ keV.
The higher the gas temperature and the observing frequency, the larger is
the deviation of the intensity change from the non-relativistic value
(Rephaeli 1995b). This must be taken into account in the analysis of high
frequency observations of individual rich clusters (Rephaeli 1995a,
Holzapfel \ea 1996). Here we are interested in the integrated effect due
to many clusters, and since we expect this to be largely dominated by the
numerous low richness (and, correspondingly, low temperature) clusters
(Colafrancesco \ea 1994), we retain here the
considerable degree of simplicity which results from expressing the full
spatial dependence linearly in the \comp parameter. The overall effect
of this simplification on our results is assessed in the Discussion.

The cumulative \comp parameter $y({\hat \gamma})$ along the line of
sight (los) $\hat \gamma$ is the sum over all clusters whose gaseous
spheres are intersected by this los. Identifying a cluster in the
ensemble by
its mass $M_m$ and redshift $z_l$, the expression for $y({\hat \gamma})$
can be written as (Cole and Kaiser 1988):
\begin{equation}
y({\hat \gamma}) = \sum_{l,m} n_{l,m} y_o(M_m, z_l)
\zeta(\vert {\hat \gamma} - {\hat \gamma}_l\vert, M_m,z_l) \, .
\end{equation}
Here $n_{l,m}$ is the occupation number of clusters in the $M-z$ space, while
\begin{equation}
y_o(M_m, z_l) \equiv (2 tg^{-1}p) (kT/m_ec^2) \sigma_T n_o r_c \cdot
\end{equation}
is the value of the \comp parameter along a los,
$\hat{\gamma_l}$,
through the center of the cluster (with mass $M_m$ at redshift $z_l$).
The angular profile $\zeta$, obtained by integrating the truncated
profile of Equation (6) along different directions is
\begin{equation}
\zeta({|\hat {\gamma}-\hat {\gamma}_l|,M,z})=
tg^{-1}
\bigg[ p \sqrt{1 - (\theta/p\theta_c)^2 \over 1 + (\theta/\theta_c)^2} \bigg]
\bigg/\biggl[(tg^{-1} p) \cdot \sqrt{1+({\theta /\theta_c})^2}\biggr]
\end{equation}
where ${\rm cos}(\theta)= \hat {\gamma} \cdot \hat {\gamma}_l$,
$\theta \leq p \theta_c$, and $\theta_c =\theta_c (M, z)$ is the
angle subtended by $r_c$.

The distribution of IC gas has characteristic angular scales reflecting
the apparent sizes of clusters at various redshifts (unlike the
case of a possible uniform intergalactic medium). Nevertheless, even
though the distribution of values of the \comp parameter
depends on the beam size, its true sky-averaged value is beam-independent.
To find the mean value of the Comptonization parameter, we average $y({\hat
\gamma})$ first over the sky and then over the ensemble of cluster
distributions.
The sky and the ensemble average act on the profile and the phase
space density respectively, yielding
\begin{equation}
\overline{\zeta(M,z)}= {\theta_c^2 \over 2 tg^{-1} p} (p - tg^{-1}p) .
\end{equation}

%\ch {\bf No need for the deleted expression.}
% and $\langle n_{l,m} \rangle = N(M,z) dM dV$.
Altogether, the mean \comp parameter averaged over the ensemble of clusters
and over the sky is
\begin{equation}
\overline{y}= \int {dV\over dz}dz \int N(M,z) y_o(M,z) \overline
{\zeta(M,z)} dM \, .
\end{equation}

In Table 2 we list values of $\overline{y}$, in the selected open
and flat models with the
parameters of Table 1
and $M_{min}=10^{13} \msun$.
The values of $\overline{y}$ are generally well below the
FIRAS upper limit, $1.5\times 10^{-5}$ (Fixsen \ea 1996).

Because values of most of the parameters are poorly known, it is
important to consider the full parameter range.
In Table 2 we also list predicted values of $\overline {y}$ obtained
with no gas evolution (\ie $\eta=0$, $\xi=0$), and with different choices
for $M_{min}$. Based on these estimates we
conclude that the assumed degree of IC gas evolution reduces
$\overline {y}$ by a factor of few, except in the BDM models
where this factor can be very significant. The decrease in the
degree of \comp with increasing $M_{min}$ is dramatic only in the
BDM $\Omega_0=0.1$ model. The predicted value of $\overline {y}$
in the BDM $\Omega_0=0.1$ model with no gas evolution is inconsistent
with the current FIRAS limit, even if $M_{min}$ is increased by a
factor of $\sim 3$ to $M_{min}=10^{13} \msun$.
%as well as the CDM + \Lambda and open CDM

As shown in Figure 1, most of the contribution to the mean \comp comes
from clusters at $z \simlt 2$, if there is gas evolution. For open CDM
models without gas evolution there is a substantial contribution to
$\overline{y}$ up to $z \simlt 5$.

Another useful quantity which can be readily calculated using the
above formalism is the mean optical thickness to Compton scattering due
to gas in clusters. Since the optical thickness of a cluster is
$\t = y \cdot m_ec^2/kT$, the above expressions can be used to calculate
$\overline{\t}$ by simply replacing $y_o(M,z)$ in Equation (15)
with $\t_o(M,z)\equiv y_o(M,z)m_ec^2/kT$,
the optical thickness along a los through the center of a cluster of
mass $M$ at redshift $z$.
Values of the mean sky-averaged optical thickness of gas in clusters
are listed in Table 2. These give a measure of the minimal degree of
scattering of the radiation from distant sources.
The sensitivity to $M_{min}$ is larger in this case: nonetheless, values
$\simlt 10^{-4}$ are obtained for most of the models. Only in the BDM
models with $n=-1$ and without gas evolution
does $\t$ attain values of $\simgt 10^{-3}$.
\bigskip

\vskip 0.3truecm
\noindent
{\Large\bf 4 \, Cluster Counts}

\smallskip
\noindent
With respect to the incident radiation field, the change of the CMB
intensity across a cluster can be viewed as a net flux emanating from
the cluster. The flux is negative below the crossover frequency and
positive above this characteristic frequency ($\simeq 217$ GHz in the
nonrelativistic limit). While the main emphasis so far has been the
measurement of the S-Z effect in individual clusters, the capability
to observe a large number of clusters in a satellite survey enhances
interest in S-Z number counts. The number of clusters observable in
sub-mm bands is of particular interest in view of forthcoming experiments
dedicated to the study of the small scale structure of CMB anisotropy.
In the nonrelativistic limit, the change of spectral intensity across
a cluster of a given $y$, observed at a frequency
$x=h\nu / kT_{CMB}$ [$T_{CMB} =(2.726 \pm 0.010) $ K; Mather \ea 1994),
is
\begin{equation}
\Delta I_\nu = {2 (kT_{CMB})^3 \over (hc)^2} g(x) y \, ,
\end{equation}
where
\begin{equation}
g(x)={x^4 e^x \over (e^x -1)^2 } \cdot [x coth(x/2) -4] \, .
\end{equation}

The differential flux measured at a given frequency from the cluster is
\begin{equation}
\Delta F_\nu (\hat{ \gamma_l}) =
\int_{4 \pi} d\Omega R_s({\vert \hat \gamma - \hat{\gamma_l} \vert }, \sigma_B)
\Delta I_\nu(\hat {\gamma}) \, ,
\end{equation}
where $R_s$ is the angular response of the receiver, say with a
Gaussian beam of dispersion $\sigma_B$, whose axis coincides with the los
to the cluster center.
The angular dependence of $\Delta I_\nu$ is
fully contained in the cluster profile, and so $\Delta F_\nu$ can be
computed by convolving this profile with the response of the receiver.
From Equations (17) and (19) the beam convolved cluster signal is:
\begin{equation}
\Delta F_\nu = {2 (kT_{CMB})^3 \over (hc)^2} g(x) y_o(M,z)  \Xi(M,z) \, ,
\end{equation}
where
\begin{equation}
\Xi \equiv \int d\Omega R_s({\vert \hat \gamma - \hat \gamma_l \vert })
\zeta({\vert  \hat \gamma_l\vert, M,z }) \, .
\end{equation}

The receiver measures the flux integrated over its passband $E(\nu)$:
\begin{equation}
\overline{\Delta F_\nu }= { \int d\nu \Delta F_\nu
E(\nu) \over \int d\nu E(\nu)} \, ,
\end{equation}
where $E(\nu)$ is the frequency response of the instrument.
Thus, the predicted number of clusters with a net flux
$>\overline {\Delta F_\nu}$ is
\begin{equation}
N(>\overline {\Delta F_\nu}) = \int {dV \over dz} dz \int_{\overline {M}
(\overline {\Delta F_\nu},z)} dM N(M,z) \, ,
\end{equation}
The lower bound of the mass integral, $\overline{M}$, is determined
from the requirement that the source flux is $> {\overline {\Delta F_{\nu}}}$.
Equation (20) provides the mass dependence of the flux,
$\Delta F_{\nu} (M)$, a monotonic function of the mass at a given
$z$.
%(see Figure 2).

In our specific estimates of the predicted number counts for the
C/S experiment, we take the spectral response to be uniform over
the two passbands centered on $140$ and $400$ GHz with widths
$\Delta \nu / \nu = \Delta x / x$ $= 0.4$ and $ 0.7$, respectively.
In Figures 3 and 4 we show the cluster number counts at the C/S's 140 and 400
GHz bands
in the different models, with and without IC gas evolution.
In Figure 5 we compare the counts expected in few models for the 400 GHz
channel.

The receivers of the C/S HFI and BOOMERANG
are bolometers with a Noise Equivalent Power (NEP) of
$\approx 10^{-17} ~W Hz^{-1/2}$. The limiting flux is expected to be
\begin{equation}
{\overline F}^{noise}_\nu \approx { NEP\over \Delta \nu \sqrt{t}}
{1\over A\epsilon} \, ,
\end{equation}
where t is the integration time in seconds, A is the effective  mirror area in
squared meters, and $\epsilon$ is the total (optical and electric)
efficiency of the system. For C/S, NEP values of $1.0\times 10^{-16}$ and
$2.7\times 10^{-17}~W Hz^{-1/2}$ are expected at $140$ and $400$ GHz,
respectively. The C/S telescope has an effective diameter of $1$ meter, and
total efficiency of $\approx 0.30$, so that with an
integration time of one year (assuming full sky coverage) we have
${\overline F}^{noise}_\nu \approx 160$ and $20$ mJy at $140$
and $400$ GHz, respectively. A similar calculation of the
BOOMERANG limiting flux -- with the same values of $A$ and $\epsilon$ but
with $NEP \approx 2 \cdot 10^{-17}$ W Hz$^{-1/2}$ and
$t \approx 225$ s per pixel -- yields
${\overline F}^{noise}_\nu \approx 20$ mJy.
In Tables 3 and 4 we give the number of clusters with flux greater
than ${\overline F}^{noise}_\nu$ and $3 {\overline F}^{noise}_\nu$,
corresponding to 1 and 3 sigma detection, respectively,
as expected in the various models.
Note that these numbers are obtained by requiring that the flux
collected by the receiver from the center of the  cluster is
greater than the limiting flux.
This estimate could be conservative for extended clusters if the flux
emanating from the central region  is under the detection limit.
However, a further smoothing of the observed map could help in
extracting  additional clusters out of the noise (the noise of
the smoothed map is lower by the factor
$\theta_{smooth}/\theta_{obs}$, where
$\theta_{smooth}$  and $\theta_{obs}$ are the resolutions of the
smoothed and observed maps, respectively).
This is reflected in Figures 3 and 4, where we show
the cluster number counts expected after smoothing the original map.
The noise level of the smoothed maps at 400 GHz is
$10.3$, $3$ and $1.5$ mJy for final resolutions of 10, 30 and 60 arc-minutes,
respectively. Correspondingly, the number of clusters that can be
detected increases.

Even under the  more conservative assumption of having just the
central pixel above the limiting flux, the predicted counts
are fairly large for C/S for which a
full sky coverage is assumed;
the quoted numbers for the BOOMERANG's 150 GHz band
($\Delta x / x =0.2$) refer to a $10^o \times
10^o$ patch of the sky. Thus, given that the S-Z effect can be identified
by virtue of its characteristic spectral signature, these experiments
can, in principle, produce S-Z catalogs of clusters.
Correlation analyses of S-Z and X-ray measurements of clusters will
provide useful information on cluster evolution, and possibly also on
cosmological scenarios.

In Figure 6 we show the redshift distribution of the  predicted number
of clusters seen above
the limiting flux at $400$ and $140$ GHz.
For flat (CDM and MDM) models $99 \%$ of the predicted clusters have
redshifts less than $z \sim 0.2$, quite independently of the  degree
of gas evolution.
In low density (either open or with cosmological constant) models
clusters are expected also at higher redshifts ($z \simlt 1$).
However, considering gas evolution in these low density models,
$\approx 90 \%$ of the  predicted clusters have $z \simlt 0.3$.

Most of the detected clusters are expected to be contained
just in one pixel of the mm and sub-mm maps, and so it will be difficult
to clearly identify clusters under these circumstances. It is therefore
important to determine also the distribution of angular sizes of clusters
whose fluxes are higher than the above limiting values. This is accomplished
by calculating the distribution
\be
N(\theta_{FWHM},z)=N(M,z) {dM \over d \theta_c} {d \theta_c \over
d\theta_{FWHM}}
\ee
where $\theta_c= D_A r_c$ is the
angle subtended by the core radius of a cluster, which
depends on $M$ through Equation (8),
$D_A$ is the angular diameter distance, and
$\theta_{FWHM}=\theta_{FWHM}(\theta_c)$
is the FWHM of the beam convolved cluster profile:
\be
\tilde{\zeta} (\theta, \sigma_B) = {1 \over \sigma_B^2} \int
d \psi \psi \zeta(\psi) exp\bigg( - {\psi^2 + \theta^2 \over 2 \sigma_B^2}
\bigg) I_0 \bigg( {\psi \theta \over \sigma_B^2} \bigg)
\ee
In Figure 7 we show the quantity
$$
{\cal N}(> \theta_{FWHM})=
\int_{\theta_{FWHM}}^{\infty} d\theta_{FWHM}'
\int dz N(\theta_{FWHM}',z) \, ,
$$
where the double integral is performed so
%by noting
that the mass of a cluster with a flux
$> {\overline F}_{\nu}^{noise}$
is larger than a minimum value $\overline {M}$ (see Equation 23);
correspondingly, at a given $z$, such a cluster has a FWHM angular
size larger than $\overline {\theta_{FWHM}}$.

\vskip 0.3truecm
\noindent
{\Large\bf 5 \, Discussion.}
\smallskip

The approach described in this paper is phenomenological, as
it is based on the normalization of the predicted cluster number
counts so they reproduce the locally observed distribution.
We consider our normalization to the XRLF to be quite robust, in the
sense that a statistically significant fit to the data is obtained over a
range of X-ray luminosities from $\sim 10^{43}$ to $\simgt 10^{45}$ erg
s$^{-1}$. The quality of this fit is low for MDM $\Omega_{\nu}=0.3$
models with scale invariant initial conditions, unless $n$ is
increased to values $\sim 1.2 \div 1.4$ (see Lucchin \ea 1996 for a
discussion).
BDM models with $n$ in the range $-1 \div 0$ substantially overproduce
the number of clusters, so the consideration of these models here is
essentially motivated by didactic purposes, as an example of pure
power law spectra.
Affecting a different normalization -- for example,
normalizing directly to the mass or temperature distributions -- can, in
principle, yield different results, due to the different  implied
shapes and amplitudes of $N(M,z)$.

A comparison of the number counts for flat and low density CDM cosmologies
shows that the counts increase with decreasing $\Omega_0$.
With our normalization to the XRLF, lowering $\Omega_0$ has the effect
of lowering $N(M,z)$ (which is proportional $\rho$) and reducing the
cutoff mass.
As a result, the local abundance of clusters in a low density universe
is lower than in the $\Omega_0=1$ case, if we assume the same values for
$h, {\cal I}$ and $b \delta_v$.
This effect is counterbalanced by lowering $b \delta_v$ and increasing
$h$ and/or ${\cal I}$ (see Table 5).
In addition, in  a low density universe $N(M,z)$ evolves less than in a
flat universe (see Fig.5). We stress that our scaling laws imply that
the core
radius is larger in low density a universe as compared to a flat
model. With all the other parameters fixed, the net flux from the
cluster is then increased.
Altogether, without gas evolution, we do expect a factor $\approx 5$
more clusters in a low
density than in a critical CDM universe. This difference is reduced
when including evolution of the IC gas
which has the effect of making the population of predicted
clusters a more local ($z \simlt 0.4$) one.
Note the strong dependence of the counts on the assumed value of the
Hubble constant.

Among the various assumptions and simplifications made in our treatment,
the uniform, spherical cluster collapse -- as described in the
Press \& Schechter (1974) formalism -- is a major one.
Another major uncertainty is the nature and degree of IC gas evolution.
The two limiting cases considered here -- no evolution, and maximum
degree of evolution which is still consistent with the EMSS distant
XRLFs -- are likely to span a reasonably realistic range. It should
also be noted that our scaling of the gas
fraction to a value of $0.1$ in a rich, local cluster may be
conservatively low, judging by some observational indications of
a value of up to 30\% (see White \ea 1993).
These uncertainties are not expected to affect our conclusion that
the sky-averaged optical depth for electron scattering in the hot IC
gas is quite low in all the models considered here. Obviously, the range
of predicted values of $\t$ constitutes the minimum level of scattering
of the CMB eversince the epoch of last scattering.
Reionization before cluster formation may have resulted in a much
higher level of scattering (Tegmark \& Silk 1995).

Another simplification made in our treatment is the use of the
non-relativistic expression for the S-Z effect. As we have noted already,
in the exact relativistic treatment the dependence of the S-Z effect is
not linear in $y$, and the the deviation of the intensity change from the
non-relativistic value is appreciable at high frequencies and gas
temperatures (Rephaeli 1995b). The major contribution to the integrated
effect due to the full cluster population comes from low temperature
clusters. The full relativistic calculation can be affected by means of
a frequency-dependent correction factor (applied to the non-relativistic
expression for the intensity change) which can be calculated at some mean,
population-weighted value of the gas temperature. Since we expect this
mean to be close to the lower end of the observed range, $3 \div 15$ keV, the
correction will generally be small except at very high frequencies. For
example, if this mean temperature is $\sim 5$ keV, then the correction
factor is a few percent in the Rayleigh-Jeans part of the spectrum, and
$\simgt 20\%$ at the $200 \div 240$ and $\simgt 600$ GHz ranges. The exact
correction factors can be calculated at each frequency and gas
temperature using the expressions given in Rephaeli (1995b).

The cluster number counts in the mm and sub-mm regions are of interest
for upcoming high-sensitivity CMB anisotropy experimentsCMB anisotropy experiments such as
BOOMERANG and C/S.
The number counts shown in Table 3 are optimistic in several ways.
First, sky confusion was not taken into account; its inclusion will
obviously reduce our predicted numbers in a way which will largely depend
on the degree of sensitivity in modeling emission from Galactic dust and
far-IR emission from other galaxies. The Galactic disk region will reduce
useful sky coverage to $\approx 80\%$. The integrated emission from galaxies
is not known, but in some models with strong luminosity or density evolution
the predicted intensity levels (e.g., Beichman and Helou 1991, Wright \ea
1994) may well exceed those corresponding to \comp. A quantitative comparison
is not warranted at this stage because of the high degree of uncertainty
in these models. In the analysis of actual data it will likely be possible
to separate out the S-Z component based on its unique spectral shape
and its larger characteristic spatial scales.
%The down-scaling of our estimated counts can be accounted for
%at the data analysis level.
For now, we include in Table 4 the
numbers of clusters with $\overline {\Delta F}_{\nu} >
3 \overline{F}_{\nu}^{noise}$. From these we can predict that if
sky-confusion is minimal, then a full sky map -- which can be
generated by C/S at 400 GHz (\ie, with a resolution of $ 4'.5$) after
one year of operation -- should include as many as $\approx 10^3$
clusters detected at the $3 \sigma$ statistical significance level.
\newpage

\begin{flushleft}
{\Large {\bf Appendix}}
\end{flushleft}

The expansion rate of a low density universe is given by the Friedmann
equations:
\be
{\dot R}^2 = {8 \pi G \rho \over 3} R^2 +
\left\{
\begin{array}{lr}
\displaystyle{ H_0^2 (1 -\Omega_0) R_0}; & \Lambda =0\\
\\
\displaystyle{ H_0^2 (1 -\Omega_0)R^2 }; & \Lambda = 1-\Omega_0
\end{array}
\right\}
\ee
The second terms in the rhs of the previous expressions describe the
curvature and the cosmological constant contributions.

Let us consider a spherical, homogeneous perturbation.
If bound, this perturbation evolves
according to the following relations:
\be
\dot {R}^2 = {8 \pi G \over 3} {\tilde \rho } R^2 +
\left\{
\begin{array}{lr}
\displaystyle{ - c^2 }; & \Lambda =0\\
\\
\displaystyle{ (1 - \Omega_0) H_0^2 R^2 - c^2}; & \Lambda = 1-\Omega_0
\end{array}
\right\}
\ee
Here $\tilde \rho$ is the perturbation overdensity and the curvature
terms ($\propto c^2$) are positive.
Note the repulsive effect  of a non vanishing
cosmological constant ($\propto 1 - \Omega_0$) on the perturbation evolution.

Let us assume that at some initial time $t_i$ the fluctuation has the
same size and expansion rate of the background.
By subtracting Equation (27) from Equation (28) we find
the conditions
\be
c^2 = {8 \pi G \over 3} \rho_i \delta_i R_i^2 +
\left\{
\begin{array}{lr}
\displaystyle{ - H_0^2 (1 - \Omega_0) R_0^2  }; & \Lambda =0\\
\\
\displaystyle{ 0  }; & \Lambda = 1-\Omega_0
\end{array}
\right\}
\ee
Here $\delta_i$ and $R_i$ are the density fluctuation and proper size
of the perturbation, $\rho_i$ is the background density at
$t_i$ and $R_m$ is the proper size of the perturbation at turnaround.

Substituting Equation (29) in Equation (28) we obtain
\be
3 R \dot {R}^2 = 8 \pi G {\tilde \rho}
R^3 - 8 \pi G \rho_i \delta_i R^2_i R  +
\left\{
\begin{array}{lr}
\displaystyle{ 3 (1 - \Omega_0) H_0^2 R R_0^2 }; & \Lambda =0\\
\\
\displaystyle{ 3 (1 - \Omega_0) H_0^2 R^3 }; & \Lambda = 1-\Omega_0
\end{array}
\right\}
\ee
Imposing the turnaround condition, $\dot R =0$, one finds in both
cases
\be
R_m^2=
\left\{
\begin{array}{lr}
\displaystyle{ {\Omega_0 \over 1 - \Omega_0} \delta_i {R_0 \over R_i}}; &
\Lambda =0\\
\\
\displaystyle{{\eta \over 2 + \eta} \bigg[ { 8 \pi G \rho_i \delta_i R_i^2
\over 3 H_0^2 (1 - \Omega_0) } \bigg]
 }; & \Lambda = 1-\Omega_0
\end{array}
\right\}
\ee
where $\eta = 3 (1- \Omega_0) H_0^2 / (4 \pi G {\tilde \rho_m})$.
Substituting Equation (31) in Equation (30) yields:
\be
3 R {\dot R}^2 =  3 H_0^2 (1 - \Omega_0) (R_m - R)\times
\left\{
\begin{array}{lr}
\displaystyle{ { 2 \over \eta } R_m^2
%\bigg(
%1 - {R \over R_m} \bigg)
}; & \Lambda =0\\
\\
\displaystyle{ (-R^2 - R_m R +{2 \over \eta} R_m^2)  }; & \Lambda = 1-\Omega_0
\end{array}
\right\}
\ee
When integrated, the previous expressions read:
\be
H_0 \sqrt{1 - \Omega_0} t =
\left\{
\begin{array}{lr}
\displaystyle{ \sqrt{\eta/2}
\int_0^x dx' \bigg[ {x' \over (1-x') } \bigg]^{1/2} }; & \Lambda =0\\
\\
\displaystyle{ \int_0^x dx' \bigg[ {x' \over (1-x')
(2/\eta - x' - x'^2) } \bigg]^{1/2}  }; & \Lambda = 1-\Omega_0
\end{array}
\right\}
\ee
where $x=R(t) / R_m$.
Analytical integration of Equation (33) with $\Lambda = 0$
provides the standard cycloid solution.

At $t_i$, by definition $x_i \equiv R(t_i)/R(t_m) \ll 1$:
this allows a Taylor expansion
of the integrands of the rhs of the previous equations and
an analytical estimate of the integrals.
One gets:
\be
\bigg( {R_m \over R_i} \bigg)^{3/2} = {2 \over 3} \sqrt {{\eta \over 2}}
{1 \over H_0 \sqrt {1 - \Omega_0 } t_i } \times
\left\{
\begin{array}{lr}
\displaystyle{ \bigg[ 1 + {3 \over 10}
\bigg( {2 \over 3} \sqrt {{\eta \over 2}}
{1 \over H_0 \sqrt {1 - \Omega_0 } t_i } \bigg)^{-2/3} \bigg]    };
& \Lambda =0\\
\\
\displaystyle{ \bigg[ 1 + {3 \over 10}
\bigg( 1 + {\eta \over 2} \bigg) \bigg( {2 \over 3} \sqrt {{\eta \over 2}}
{1 \over H_0 \sqrt {1 - \Omega_0 } t_i } \bigg)^{-2/3} \bigg]    };
& \Lambda = 1-\Omega_0
\end{array}
\right\}
\ee
We can now write the initial overdensity of the fluctuation:
\be
{\tilde \rho}_i = {\tilde \rho}_m \bigg( {R_m \over R_i} \bigg)^3 =
{1 \over 6 \pi G t_i^2} \times
\left\{
\begin{array}{lr}
\displaystyle{ \bigg[ 1 + {3 \over 5}
\bigg( {2 \over 3} \sqrt {{\eta \over 2}}
{1 \over H_0 \sqrt {1 - \Omega_0 } t_i } \bigg)^{-2/3} \bigg]
 }; & \Lambda =0\\
\\
\displaystyle{ \bigg[ 1 + {3 \over 5}
\bigg( 1 + {\eta \over 2} \bigg) \bigg( {2 \over 3} \sqrt {{\eta \over 2}}
{1 \over H_0 \sqrt {1 - \Omega_0 } t_i } \bigg)^{-2/3} \bigg]
 }; & \Lambda = 1-\Omega_0
\end{array}
\right\}
\ee
As both in the open and flat case $\rho_i= 1/(6 \pi G t_i^2)$, given
that $\tilde \rho_i= \rho_i (1 + \delta_i)$, one finally
gets:
\be
\delta_i = {3 \over 5}
\bigg( {3 \over 2} \sqrt {{2 \over \eta }}
H_0 \sqrt {1 - \Omega_0 } t_i  \bigg)^{2/3} \times
\left\{
\begin{array}{lr}
\displaystyle{ 1  }; & \Lambda =0\\
\\
\displaystyle{ \bigg( 1 + {\eta \over 2} \bigg)  }; & \Lambda = 1-\Omega_0
\end{array}
\right\}
\ee
The previous expressions give the initial density contrast (at some
arbitrary early time $t_i$) of a perturbation that will turnaround at
a given $t_m$. This information is contained in the $\eta$ parameter
through Equation (33). Substituting $t=t_m$ in eq. (31) which implies
$x=1$, we obtain a relation between the turnaround time and $\eta$.
Given the parameter $\eta$ we have (from its definition) the value
$\tilde {\rho}_m$ of the perturbation overdensity at the turnaround
time.
We assume that the perturbation eventually virializes at $t_v = 3 t_m$
and in this regime reaches a
final dimension $R_v = R_m/2$ in the open case, and $R_v = R_m (1 -
\eta/2)/(2 - \eta/2)$ in the $\Lambda=1 - \Omega_0$ case (see
Lahav \ea 1991). Having determined the virialization time, $t_v$, and the
background density at that time, we can finally evaluate the
non-linear overdensity of a virialized structure
\be
\Delta_v = { {\tilde \rho}_v \over \rho(t_v) } .
\ee
In an open universe we can write the following analytic expression:
\be
\Delta_v = {18 \pi^2 \over \Omega_0H_0^2t_v^2} {1\over (1+z_v)^3} \,,
\ee
where $t_v$ and $z_v$ are related through standard time--redshift
relations.

\newpage
\def\ref{\par\noindent\hangindent 20pt}

\noindent
{\bf References}
\smallskip

\ref Beichman, C.A. and Helou, G. 1991, ApJ, 370, L1

%\ref Bennett, D.P., Stebbins, A. and Bouchet, F.R. 1992, ApJ, 399, L5

\ref Cavaliere, A., Colafrancesco, S. and Menci, N. 1993, ApJ, 415, 50

\ref Colafrancesco, S., Mazzotta, P., Rephaeli, Y. and Vittorio, N. 1994, ApJ, 433, 454

\ref Colafrancesco, S. and  Vittorio, N. 1994, ApJ, 422, 443

\ref Cole, S. and Kaiser, N. 1988, MNRAS, 233, 637

\ref David, L.P., Arnaud, K.A., Forman, W.,  \& Jones, C. 1990, ApJ, 356, 32

%\ref Davis, M., Summers, F.J. and, Schlegel, D. 1992, Nature, 359, %383

%\ref Efstathiou, G., Bond, J.R. and, White, S.D.M. 1992, MNRAS, 258, %1p

\ref Fixsen, D.J. \ea 1996, preprint

\ref Gioia, I.M., Henry, J.P., Maccacaro, T., Morris, S.L., Stocke, J.T., \& Wolter, A. 1990, ApJ, 356, L35

\ref Henry, J.P., Gioia, I.M., Maccacaro, T., Morris, S.L., Stocke, J.T., \& Wolter, A. 1992, ApJ, 386, 408

\ref Holtzmann, J. 1989, ApJS, 71, 1

\ref Holzapfel, W.L. \ea 1996, ApJ, in press

\ref Jones, C. \& Forman, W. 1992, in
{\it Clusters and Superclusters of Galaxies}, A.C. Fabian \etal eds.,
(Cambridge: Cambridge  University Press), p.

%\ref Klypin, A., Borgani, S., Holtzman, J., and Primack, J. 1995, %ApJ, 444, 1

%\ref Klypin, A., Nolthenius, R., and Primack, J. 1995, ApJ, in press

\ref Kowalski, M.P., Ulmer, M.P., Cruddace, R.G.,  \& Wood, K.S.
1984, ApJS, 56, 403

% \ref Kompaneets AS. 1957. {\it Sov. Phys. JETP} 4, 730

\ref Lahav, O., Lilje, P.B., Primack, J.R., and Rees, M.J. 1991, MNRAS, 251, 128

\ref Lange, A. \ea 1995, in ``Infrared and sub-mm Space Missions in the Coming Decade'', Space Science Review, Kluwer Acad. Pub.

\ref Lucchin, F., Colafrancesco, S., deGasperis, G., Matarrese, S.,
Mei, S., Mollerach, S., Moscardini, L., and Vittorio, N. 1995, ApJ, in press.

% \ref Makino N, Suto Y. 1993. \apj 405, 1

\ref Mandolesi, R. \ea 1995, Planet. Space Sci., in press.

% \ref Markevitch M, Blumenthal GR, Forman W, Jones C, Sunyaev RA.
%     1991. \apjl 378, L33
%
% \ref Markevitch M, Blumenthal GR, Forman W, Jones C, Sunyaev RA.
%     \ea 1992. \apj 395, 326

\ref Mather, J.C., Cheng, E.S., Cottingham, D.A., Eplee, R.E.,
Fixsen, D.J. \ea 1994, ApJ, 420, 439

%\ref Moscardini, L., Tormen, G., Matarrese, S., and Lucchin, F. 1995,
%ApJ, 442, 469

\ref Press, W.H. and Schechter, P. 1974, ApJ, 187, 425

% \ref Rephaeli Y. 198l. \apj 245, 35l
%
% \ref Rephaeli Y. 1993. \apj 418, 1

\ref Rephaeli, Y. 1995a, ARA\&A 33, 541

\ref Rephaeli, Y. 1995b, ApJ, 445, 33

\ref Sarazin, C.L. 1988. {\it X-Ray Emission from Clusters of Galaxies}, Cambridge: Cambridge University Press

%\ref Smoot, G.F., Bennett, C.L., Kogut, A., Wright, E.L., Aymon, J. %\ea 1992, ApJ, 396, L1

\ref Sunyaev, R.A. and Zel'dovich, Y.B. 1970, Astrophys. Sp. Sci., 7, 3

% \ref Sunyaev RA, Zeldovich YB. 1972. {\it Comm. Astrophys. Sp. Phys.} 4,
%     173

\ref Tegmark, M., \& Silk J. 1995, ApJ, 441, 458

%\ref White, S.D.M., Efstathiou, G. and Frenk, C.S. 1993, MNRAS, 262, 1023

\ref White, S.D.M., Navarro, J.F., Evrard, A.,  and Frenk, C.S. 1993,
Nature, 366, 429

% \ref White M, Scott D, Silk J. 1994. \ar 32, 319

% \ref Wright EL. 1979. \apj 232, 348

\ref Wright, E.L., Mather, J.C., Fixsen, D.J., Kogut, A., Shafer, R.A. \ea 1994, ApJ, 420, 450

% \ref Zeldovich YB, Sunyaev RA. 1969. {\it Astrophys. Space. Sci.} 4, 301

\newpage

\centerline{\large Figure captions}

\vskip 0.3truecm
\noindent
{\bf Figure 1}. ~ The dependence of ${\overline y}$ from the maximum redshift,
$z_{max}$, of integration [cf. Equation (16)].
A minimum mass cutoff $M_{min} = 10^{12.5} \msunh$ has been chosen,
corresponding to the smallest systems we consider here.
Continuous and dotted curves refer to cases with and without
IC gas evolution.

\vskip 0.3truecm
\noindent
{\bf Figure 2}. ~ The integration domain of $N(> \overline {\Delta F}_{\nu})$
is shown for the case of a standard CDM model ($\Omega_0=1$, $h=0.5$,
$n=1$) and for two different channels of the C/S experiment.
The dotted contours in the Log$M$-Log$z$ plane represent isocontours
of the quantity $N(M,z) \cdot M \cdot z$ and are
spaced by a factor of $10$, the outer contour corresponding to unity.
The region contributing to the source counts is to the
right of the lines $\overline {\Delta F}_{\nu} =161$
and $19.7$ mJy
for the C/S's 140 and 400 Ghz channels, respectively.
Heavy continuous and dashed lines refer to the cases with and without
IC gas evolution.

\vskip 0.3truecm
\noindent
{\bf Figure 3}. ~ Cluster number counts
for the C/S 400 GHz channel for different angular resolutions
($4.5 $ arcmin: continuos lines; $10$ arcmin: dotted lines; $30$ arcmin:
short-dashed lines; $60$ arcmin: long-dashed lines) in different models with
(Figure 3a) and without (Figure 3b)
IC gas evolution.

\vskip 0.3truecm
\noindent
{\bf Figure 4}. ~ Same as if Figure 3 but for the C/S 140 GHz channel.
In this figure we do not plot the predictions for the $4.5$ arcmin resolution.

\vskip 0.3truecm
\noindent
{\bf Figure 5}. ~ Comparison of the cluster number counts predicted in
different models of structure formation (MDM: continuous line; open CDM: dotted
line; low density, vacuum dominated CDM: short-dashed lines; standard CDM:
long-dashed lines) for the C/S 400 GHz channel with (panel a) and without (panel b) IC gas evolution.

\vskip 0.3truecm
\noindent
{\bf Figure 6}. ~ The redshift distribution $N(>z)$ of the cluster number counts in
different models (MDM: continuous line; open CDM: dotted line; low density,
vacuum dominated CDM: short-dashed lines; standard CDM: long-dashed lines) for different sensitivities:
$\overline {\Delta F}_{\nu} =10$ and $20$ mJy (Figure 6a) and
$\overline {\Delta F}_{\nu} =30$ and $60$ mJy (Figure 6b) for the
C/S 400 GHz ($\sigma_{FWHM}=4.5$ arcmin) and 140 GHz
($\sigma_{FWHM}=10$ arcmin) channels.
In each figure panel a) and b) refer to predictions without IC gas evolution.

\vskip 0.3truecm
\noindent
{\bf Figure 7}. ~ The distribution ${\cal N}(>\theta_{FWHM})$
of the typical angular dimensions of
the clusters above the limiting flux
for the C/S 400 (Figure 7a) and
140 (Figure 7b) GHz channels.
Thin lines, with (continuous) and without (dotted) IC gas evolution,
refer to the distribution of intrinsic dimension.
Thick lines, 
with (continuous) and without (dotted) IC gas evolution, 
refer to $4.5$ and $10$ arcmin smoothing of the C/S
maps.

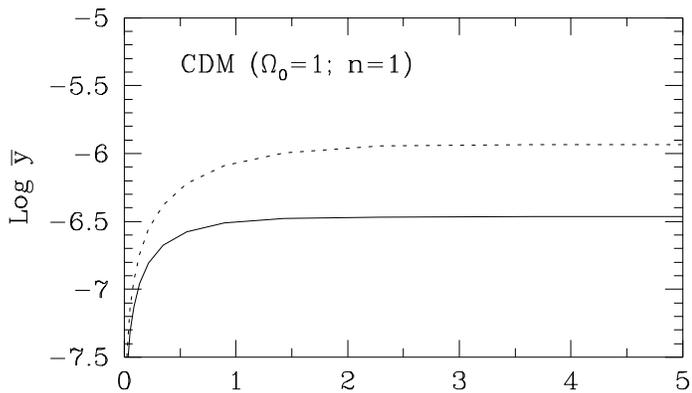
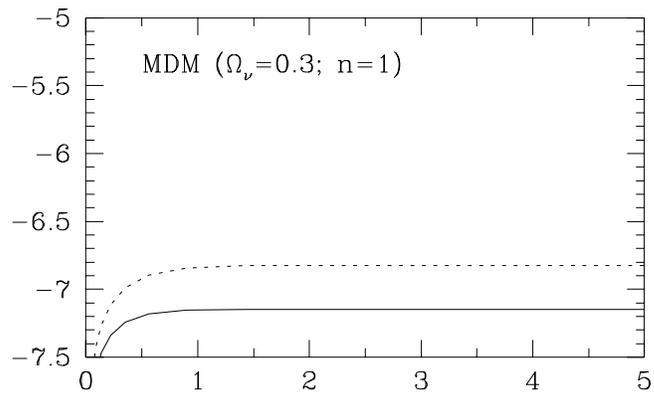
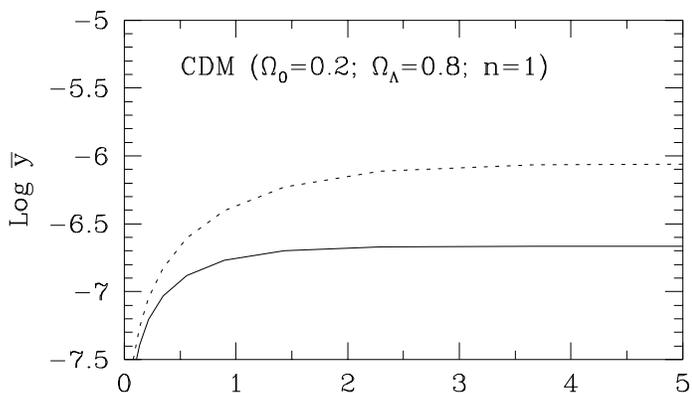
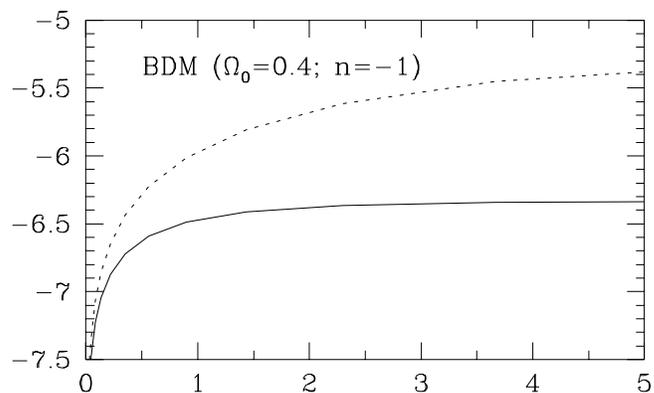
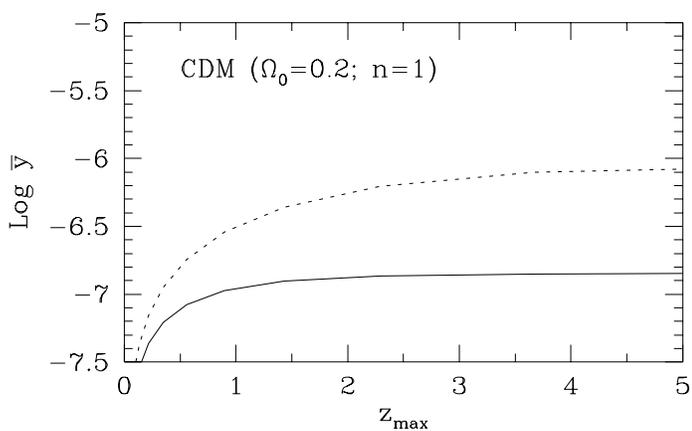
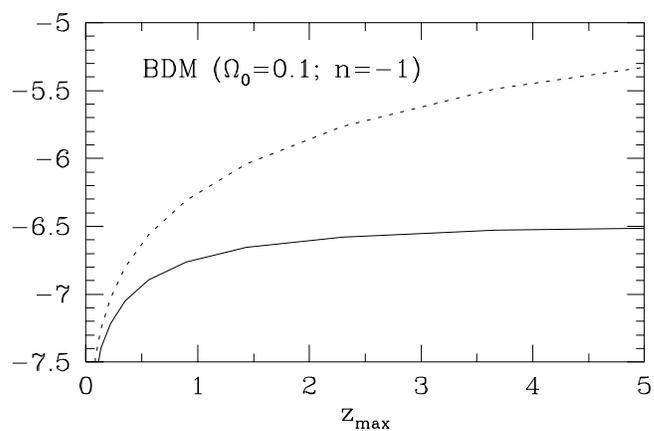
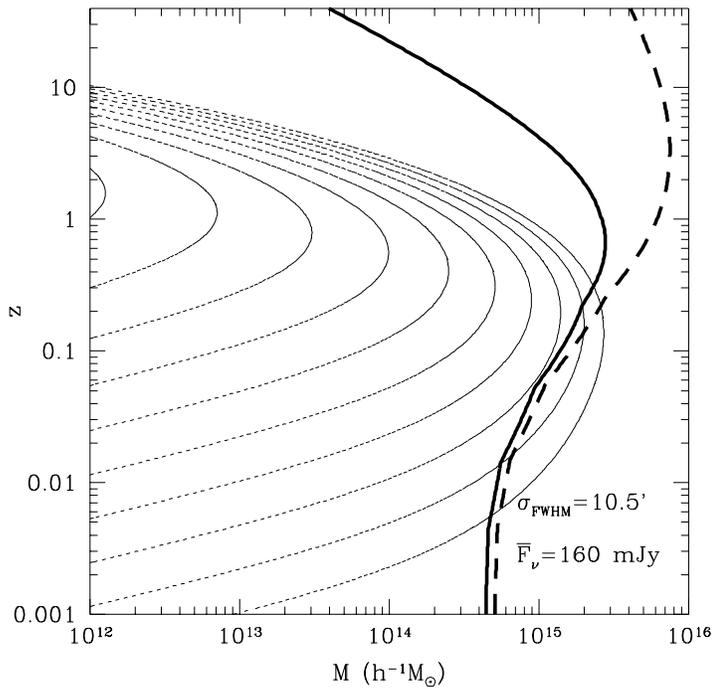
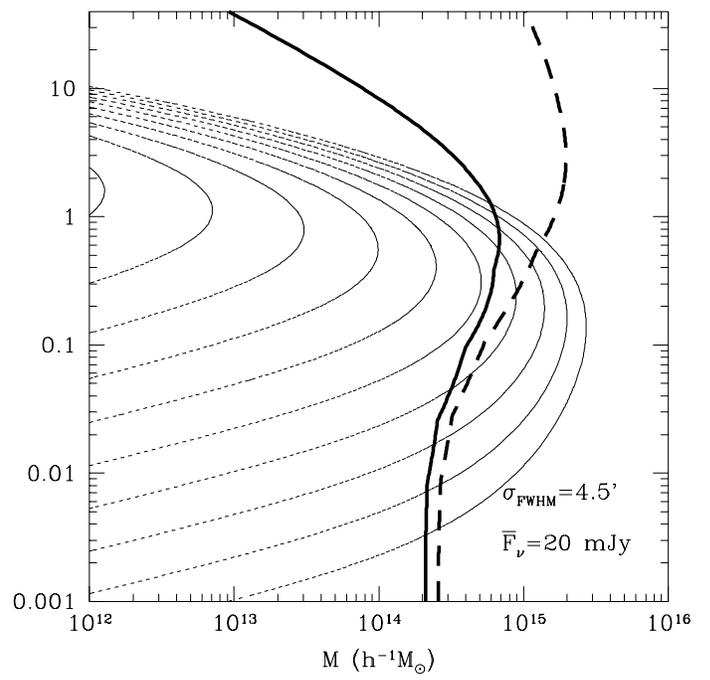
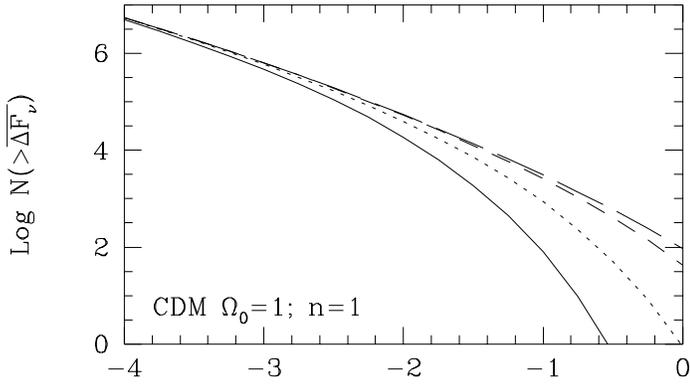
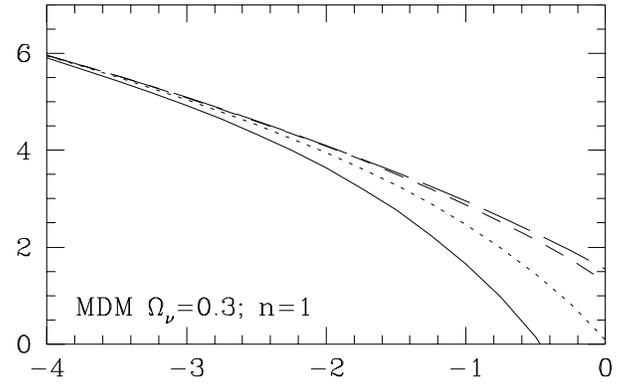
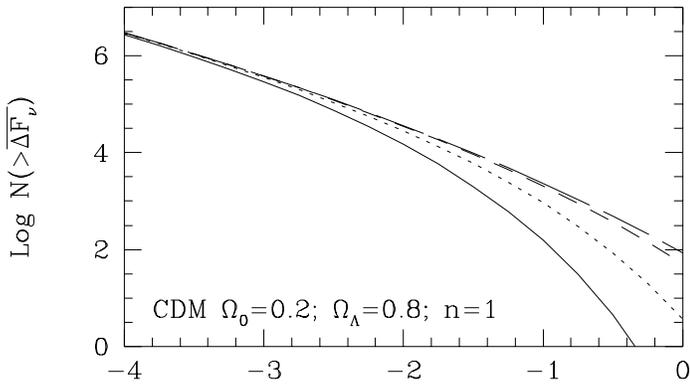
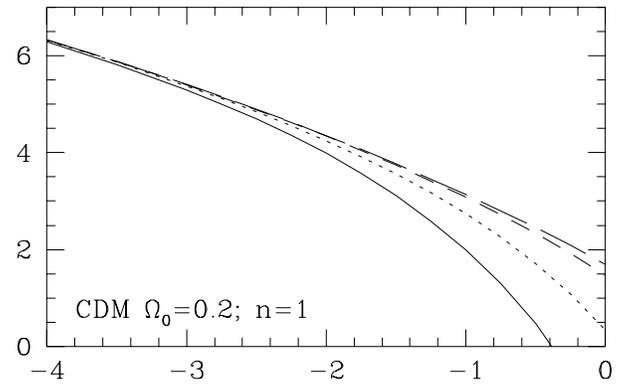
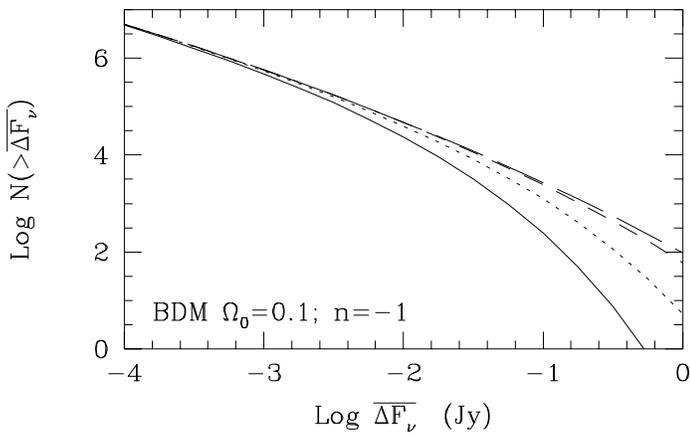
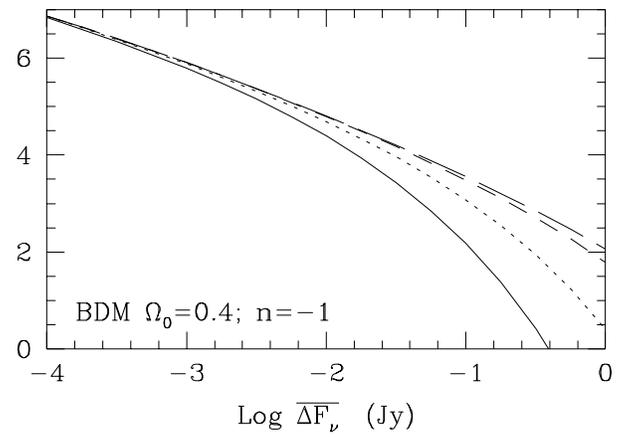
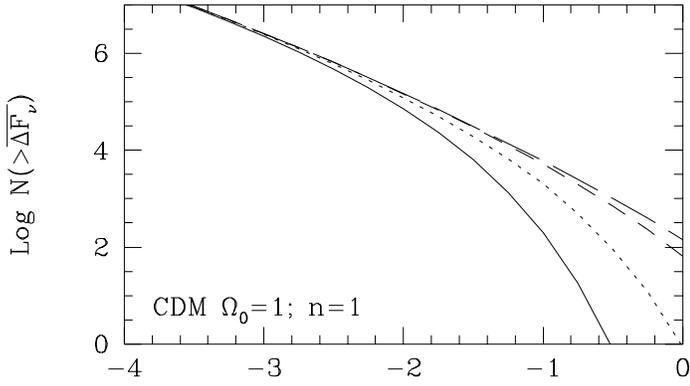
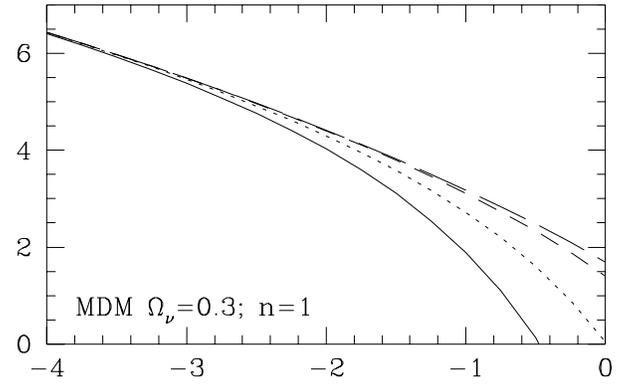
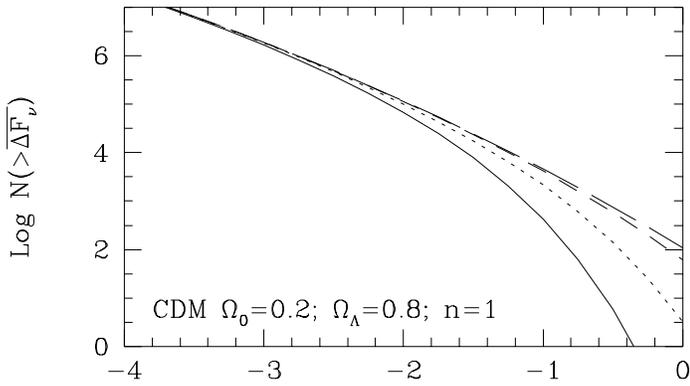
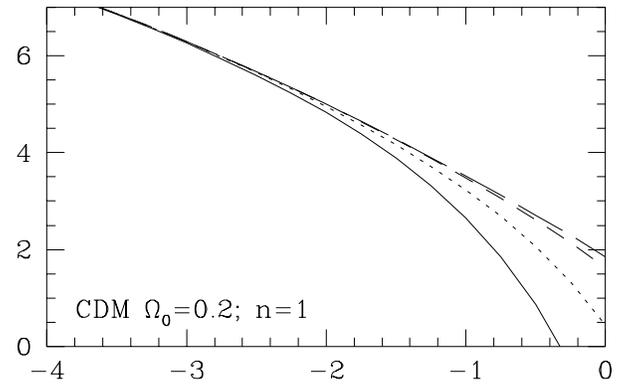
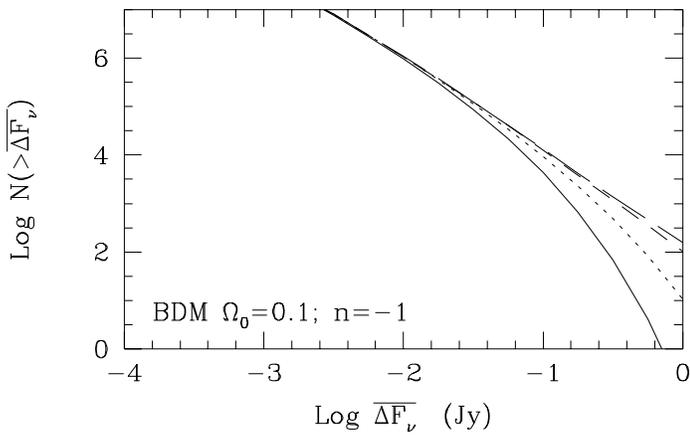
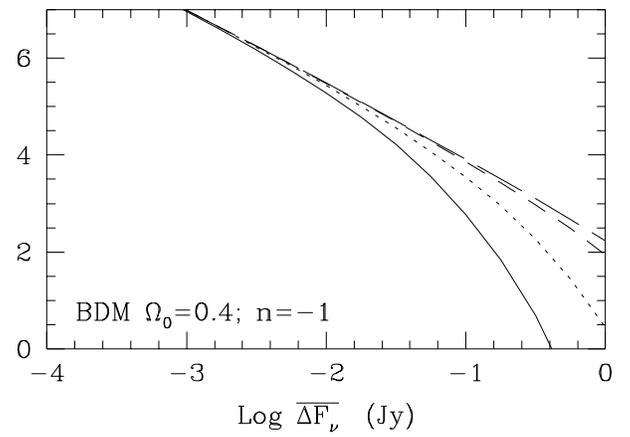
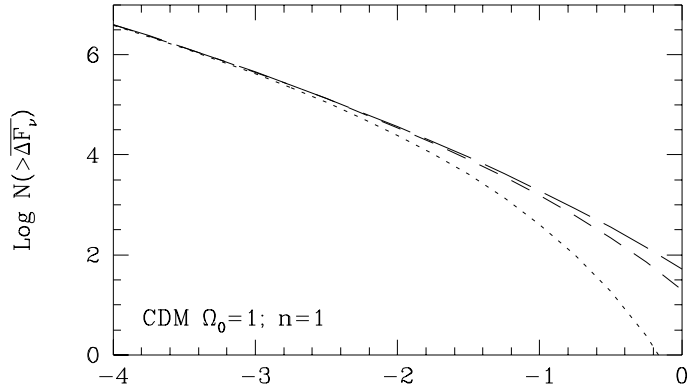
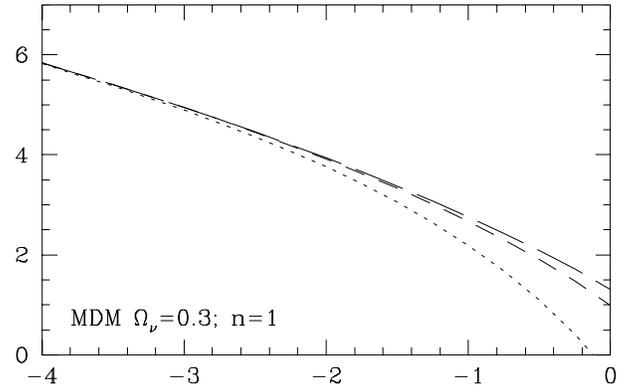
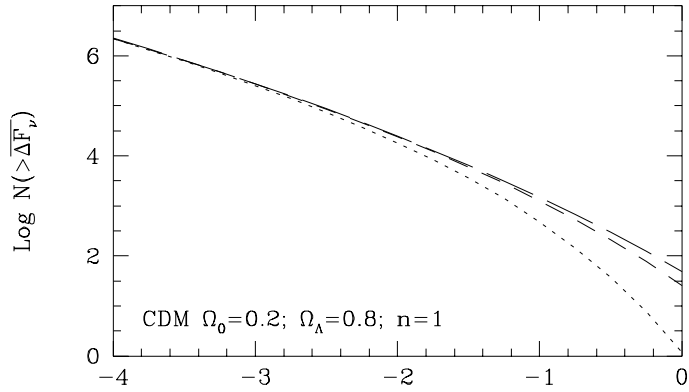
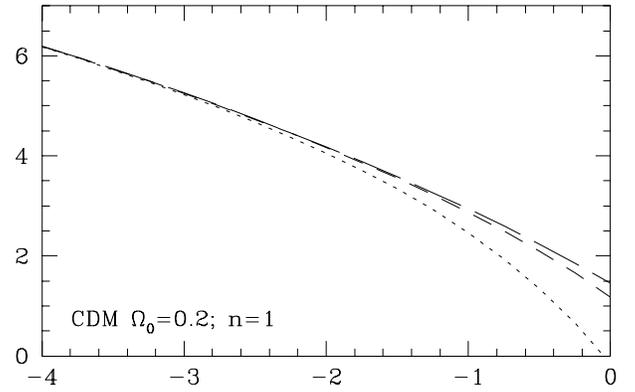
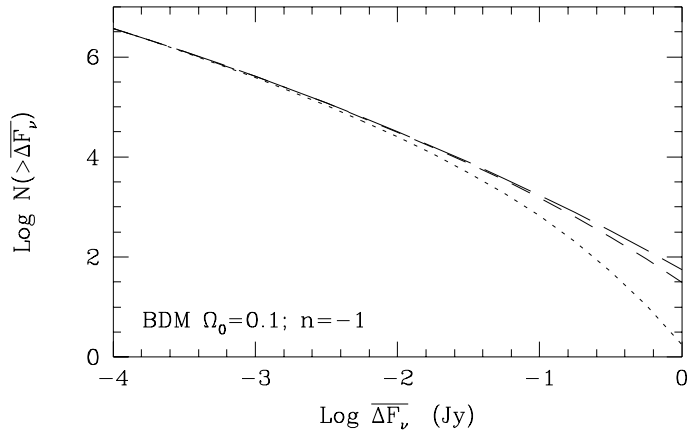
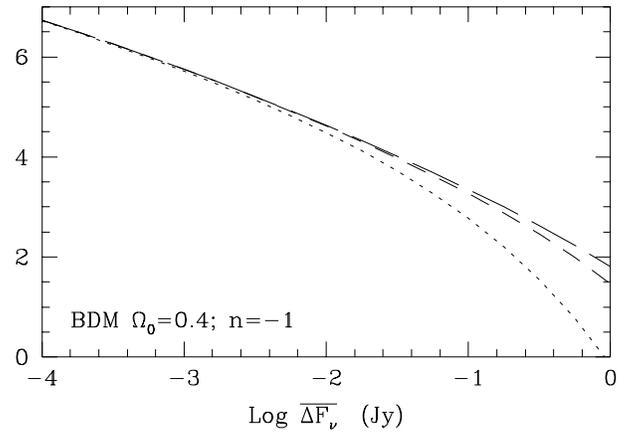
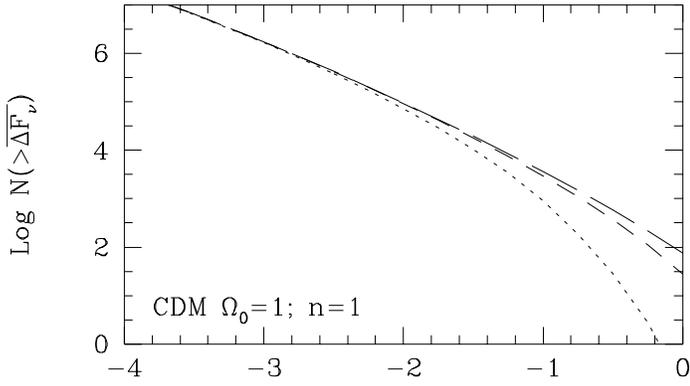
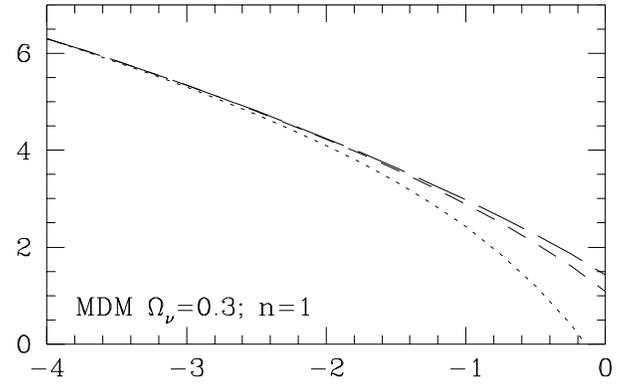
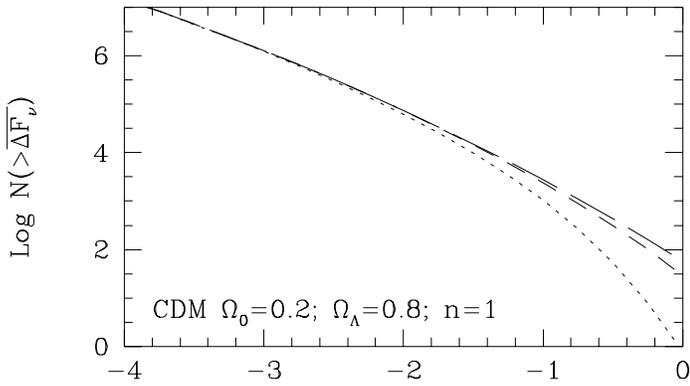
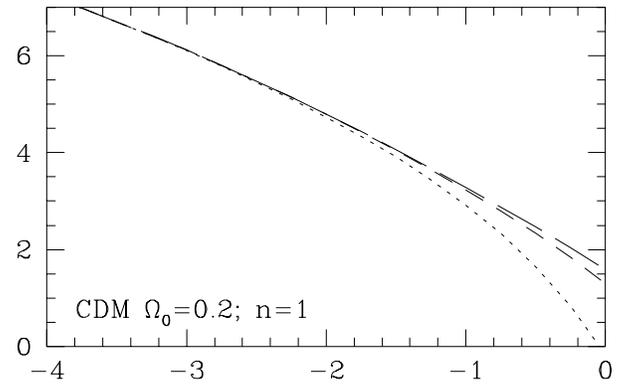
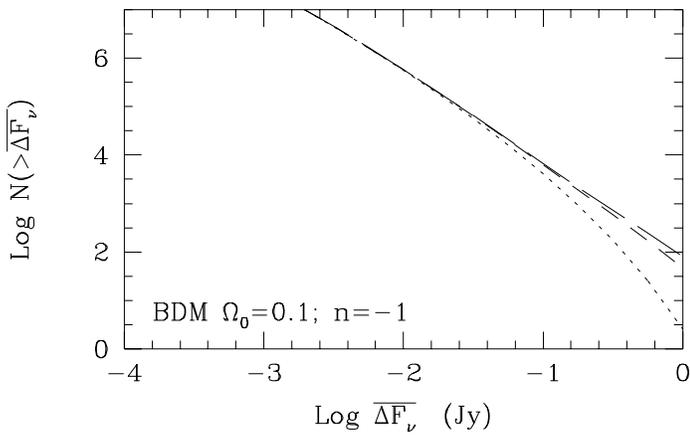
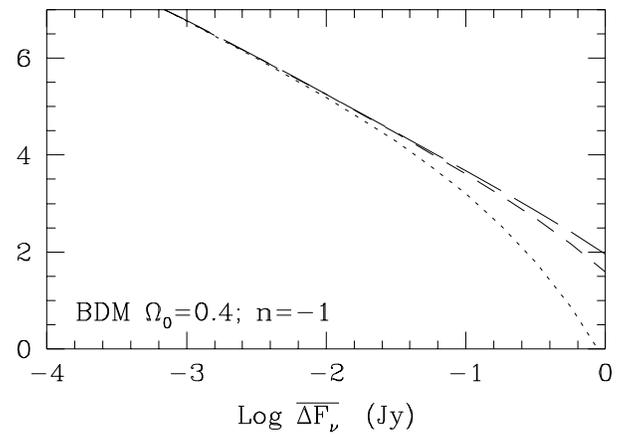
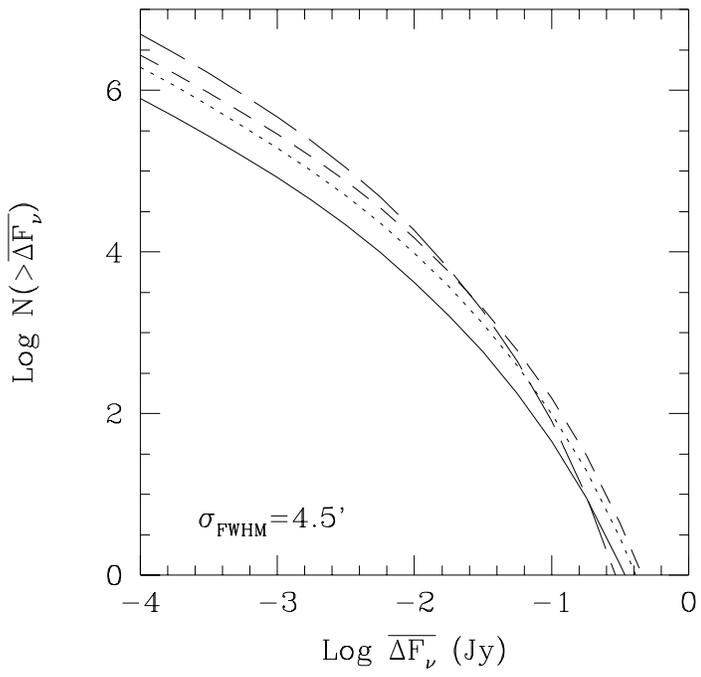
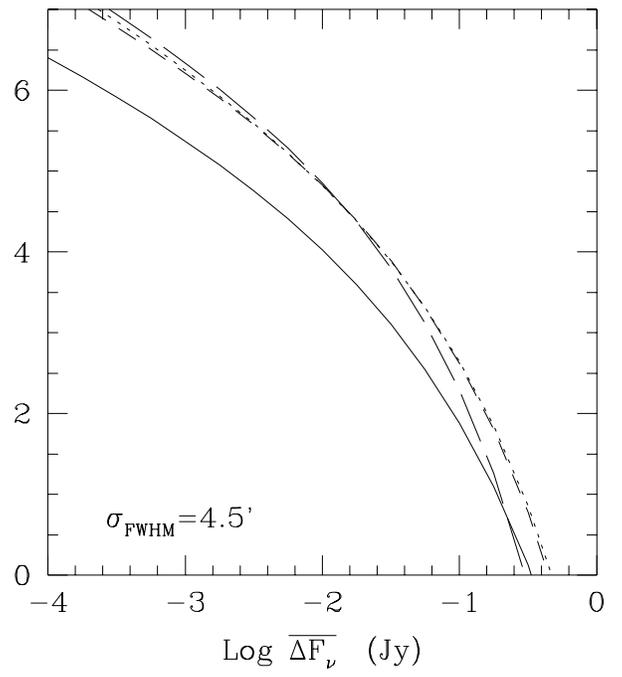
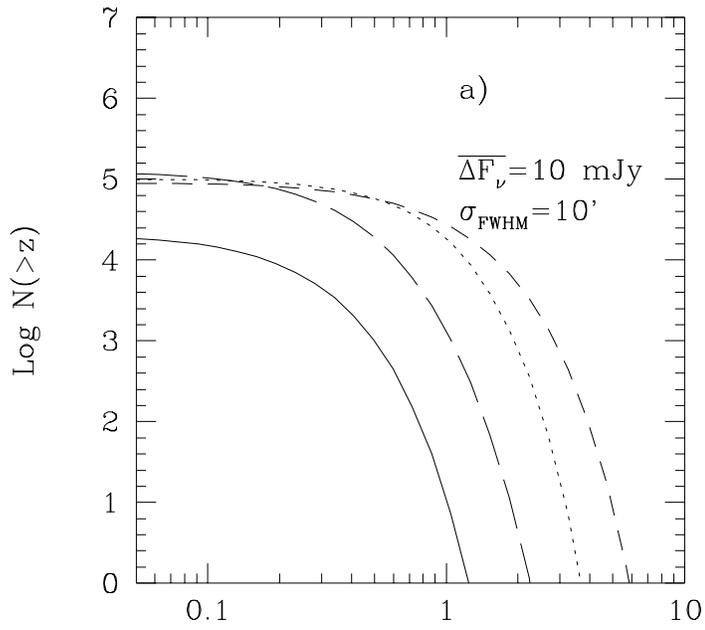
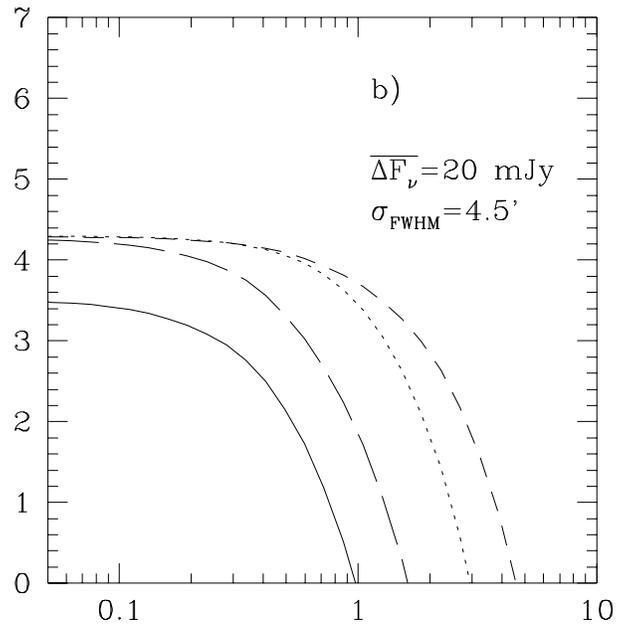
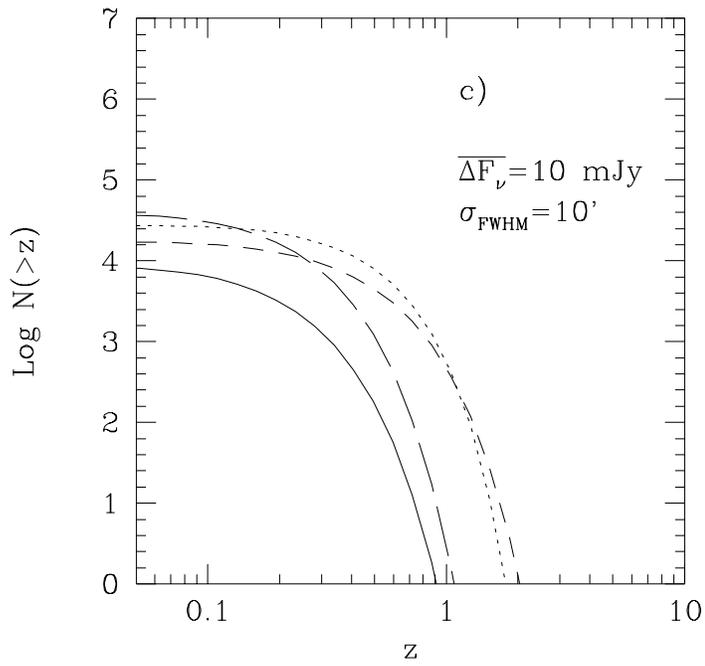
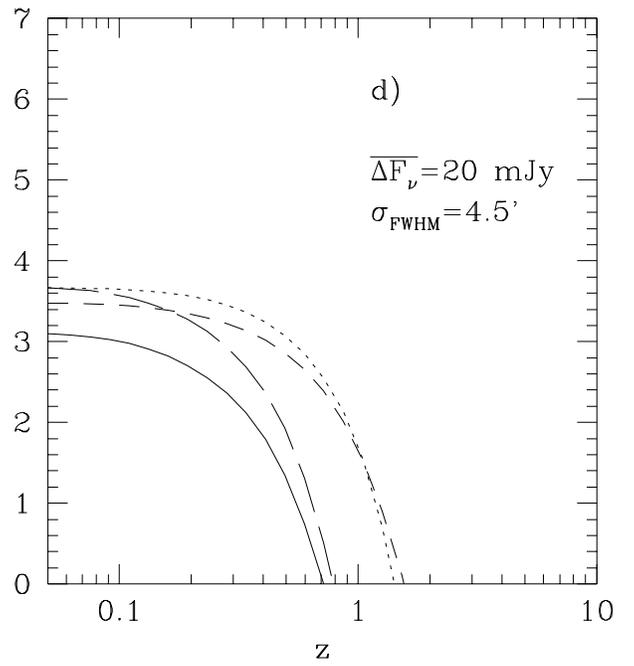
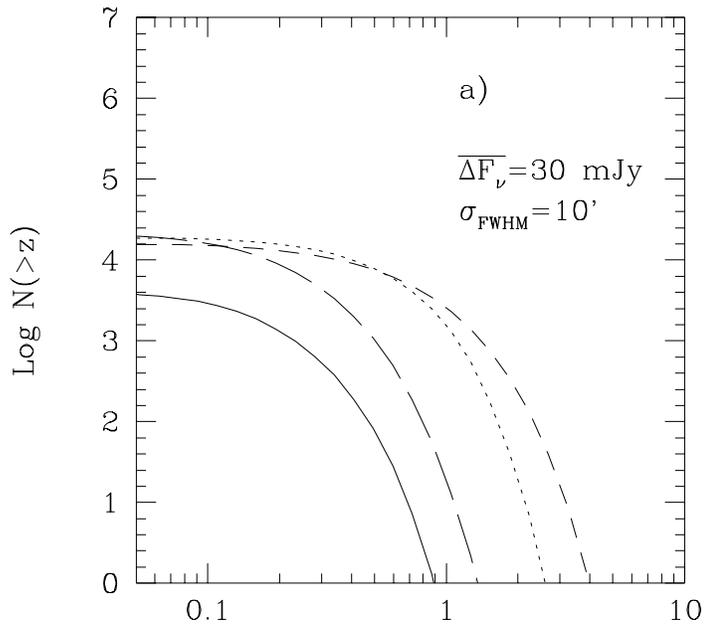
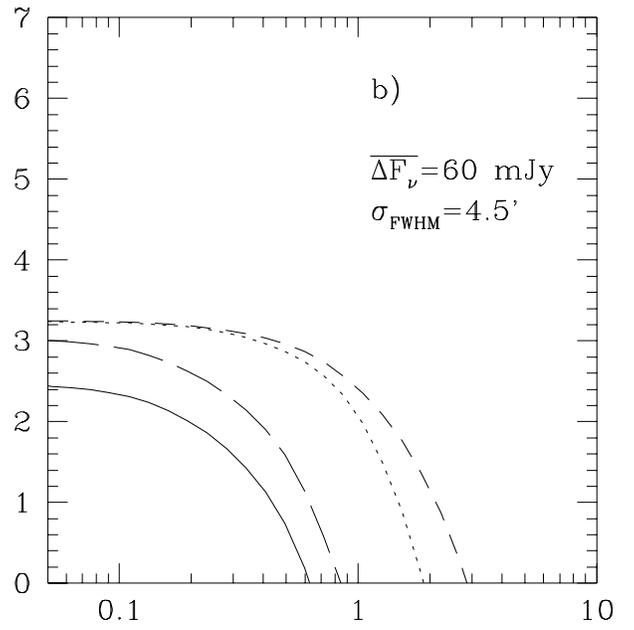
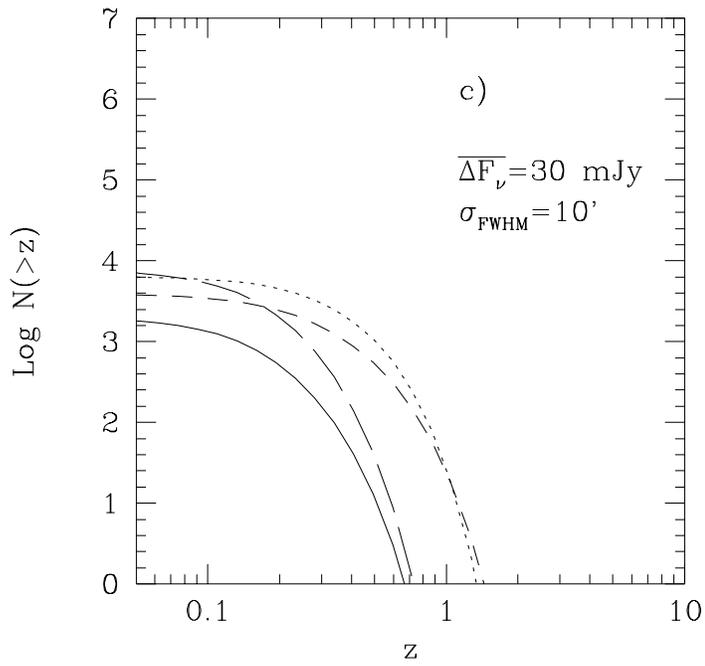
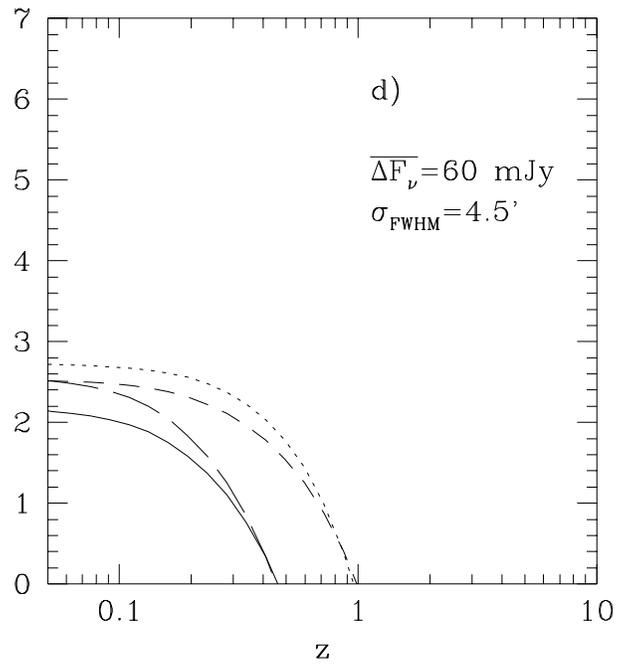
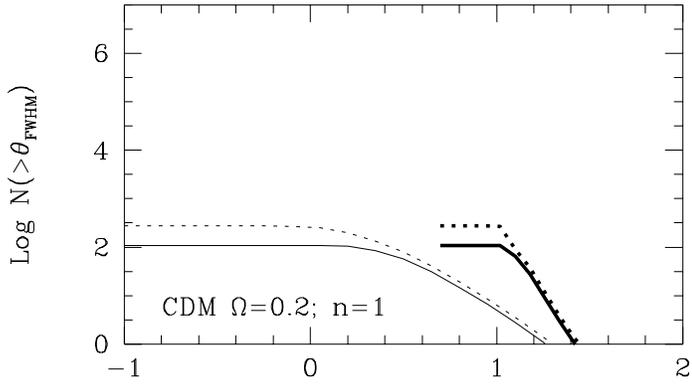
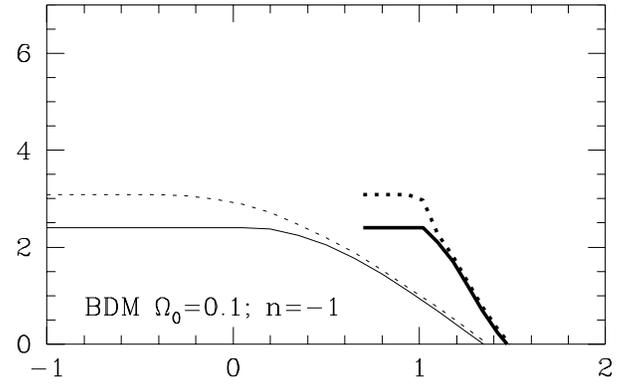
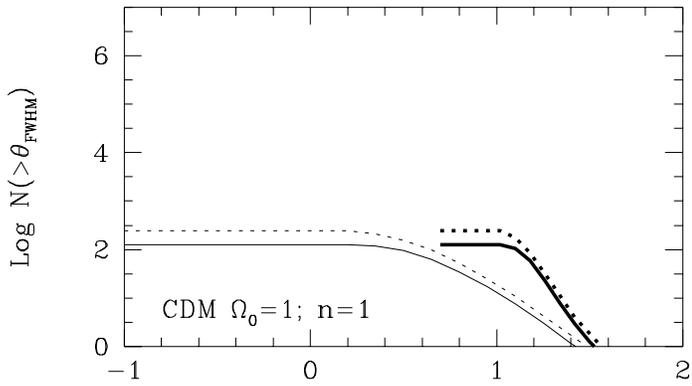
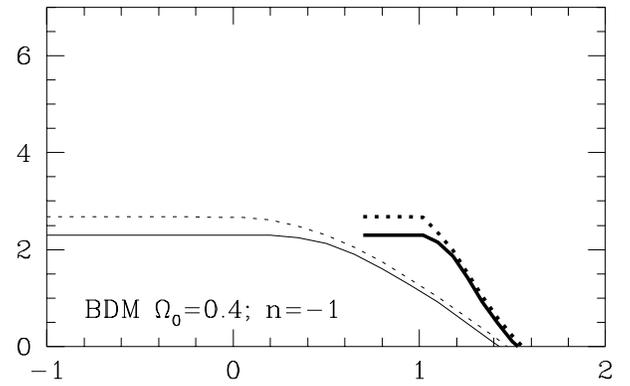
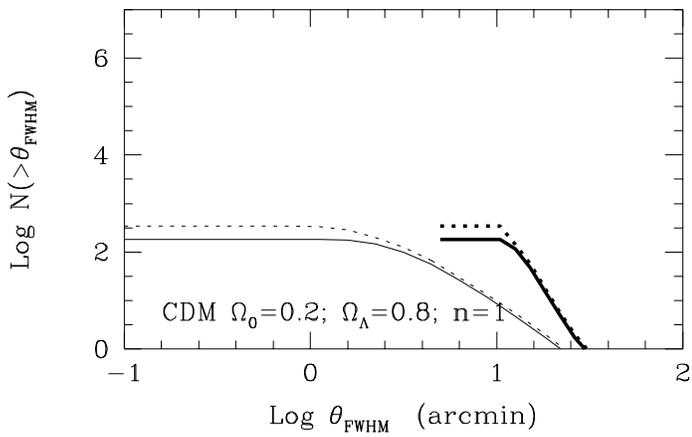
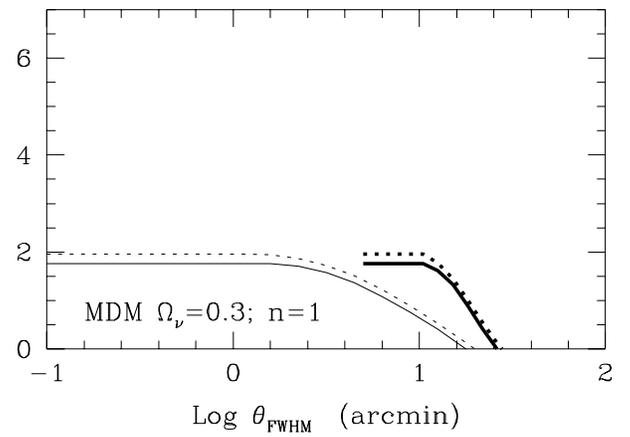
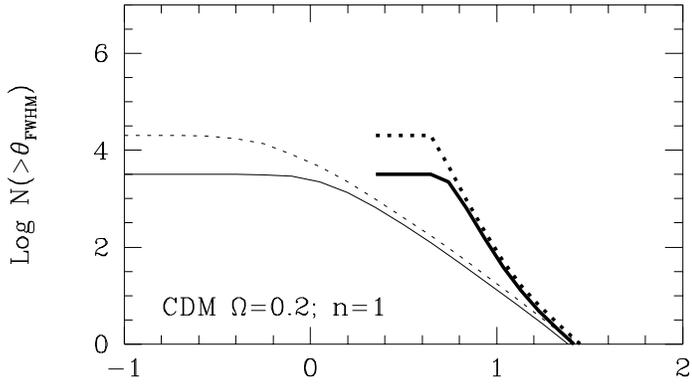
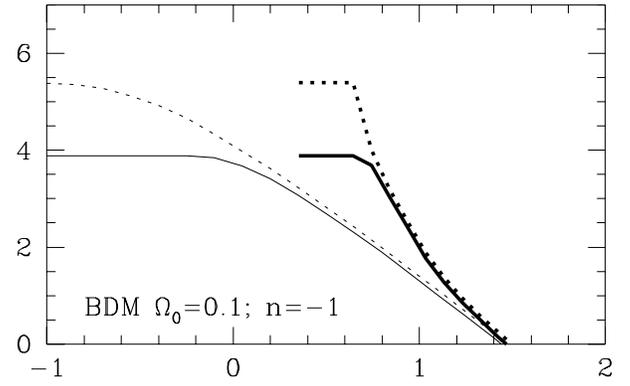
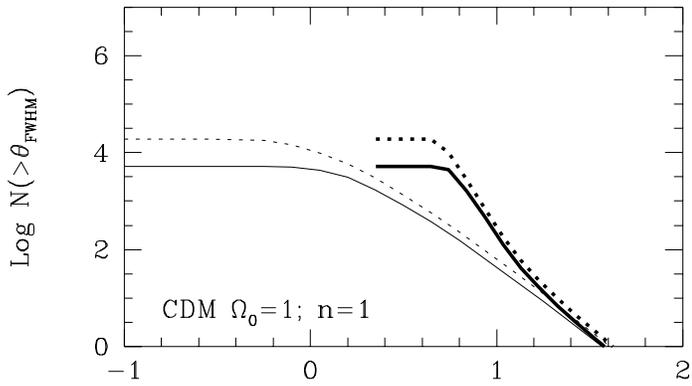
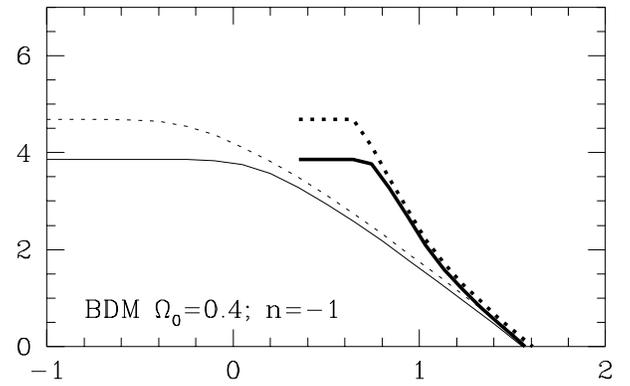
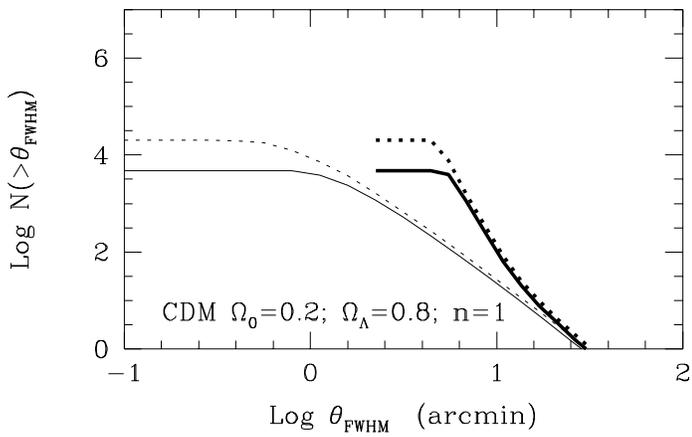
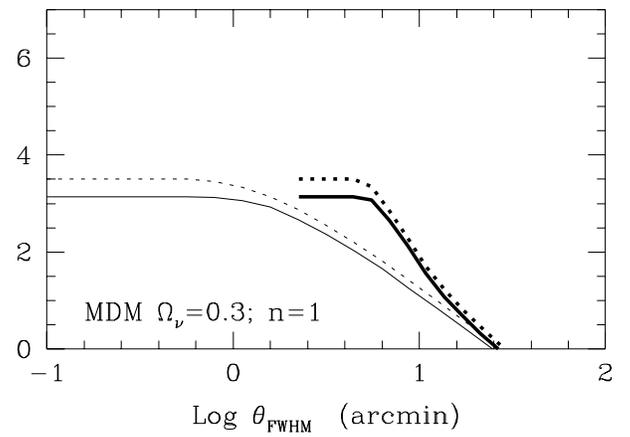
\end{document}